\documentstyle[onecolumn]{mn}
\input{epsf}

\newif\ifAMStwofonts
\AMStwofontstrue

\newcommand{\aap}{A\&A}
\newcommand{\aapr}{A\&A Rev.}
\newcommand{\aj}{AJ}
\newcommand{\apj}{ApJ}
\newcommand{\apjl}{ApJL}
\newcommand{\apjs}{ApJS}
\newcommand{\apss}{Ap\&SS}
\newcommand{\mnras}{MNRAS}
\newcommand{\nat}{Nature}
\newcommand{\pasp}{PASP}
\newcommand{\physrep}{Phys. Rep.}

\newcommand{\prd}{Phys. Rev. D}
\newcommand{\pre}{Phys. Rev. E}

\newcommand{\go}{\mathrel{\raise.3ex\hbox{$>$}\mkern-14mu
             \lower0.6ex\hbox{$\sim$}}}
\newcommand{\lo}{\mathrel{\raise.3ex\hbox{$<$}\mkern-14mu
             \lower0.6ex\hbox{$\sim$}}}
\newcommand{\be}{\begin{equation}}
\newcommand{\ee}{\end{equation}}
\newcommand{\ba}{\begin{eqnarray}}
\newcommand{\ea}{\end{eqnarray}}
\newcommand{\lp}{\left(}
\newcommand{\rp}{\right)}
\newcommand{\lb}{\left[}
\newcommand{\rb}{\right]}
\newcommand{\etal}{et al.}

\newcommand{\mpr}{m_{\rm p}}
\newcommand{\me}{m_{\rm e}}
\newcommand{\Ye}{Y_{\rm e}}
\newcommand{\sigth}{\sigma_{\rm T}}
\newcommand{\sigsb}{\sigma_{\rm SB}}
\newcommand{\alphaf}{\alpha_{\rm F}}
\newcommand{\lambdae}{\lambda_{\rm e}}
\newcommand{\vecE}{{\bf E}}
\newcommand{\vecB}{{\bf B}}
\newcommand{\veck}{{\bf k}}
\newcommand{\vece}{{\bf e}}
\newcommand{\vecr}{{\bf r}}
\newcommand{\ktot}{\kappa^{\rm tot}}

\newcommand{\kabs}{\kappa^{\rm abs}}
\newcommand{\ksc}{\kappa^{\rm sc}}

\newcommand{\krtot}{\kappa^{\rm R}}
\newcommand{\kj}{\kappa^{\rm J}}
\newcommand{\kp}{\kappa^{\rm P}}
\newcommand{\kf}{\kappa^{\rm F}}
\newcommand{\ktotj}{\kappa^{\rm tot}_j}
\newcommand{\kffj}{\kappa^{\rm ff,e}_j}
\newcommand{\kiffj}{\kappa^{\rm ff,i}_j}
\newcommand{\kabsj}{\kappa^{\rm abs}_j}
\newcommand{\kscj}{\kappa^{\rm sc}_j}
\newcommand{\kesj}{\kappa^{\rm es}_j}
\newcommand{\kesji}{\kappa^{\rm es}_{ji}}
\newcommand{\kisji}{\kappa^{\rm is}_{ji}}
\newcommand{\kisj}{\kappa^{\rm is}_j}
\newcommand{\keso}{\kappa^{\rm es}_0}

\newcommand{\kscji}{\kappa^{\rm sc}_{ji}}
\newcommand{\Kabsj}{K^{\rm abs}_j}
\newcommand{\Kscj}{K^{\rm sc}_j}

\newcommand{\Kscji}{K^{\rm sc}_{ji}}
\newcommand{\Kexc}{K^{\rm sc}_{21}}
\newcommand{\Ktotj}{K^{\rm tot}_j}
\newcommand{\mfpo}{l_0}
\newcommand{\mfpj}{l_j}

\newcommand{\taur}{\tau_{\rm R}}
\newcommand{\gff}{\bar{g}^{\rm ff}}
\newcommand{\nel}{n_{\rm e}}
\newcommand{\nion}{n_{\rm i}}
\newcommand{\Teff}{T_{\rm eff}}
\newcommand{\Inu}{I_\nu}
\newcommand{\Bnu}{B_\nu}
\newcommand{\Snu}{S_\nu}
\newcommand{\Fnu}{F_\nu}

\newcommand{\vFnu}{{\bf F}_\nu}
\newcommand{\vecF}{{\bf F}}
\newcommand{\unu}{u_\nu}
\newcommand{\unuj}{u_\nu^j}
\newcommand{\unup}{u_\nu^{\rm P}}
\newcommand{\upl}{u^{\rm P}}
\newcommand{\meani}{i_\nu^j}

\newcommand{\meanf}{f_\nu^j}

\newcommand{\omegab}{\omega_{Be}}
\newcommand{\Ebi}{E_{Bi}}
\newcommand{\omegabi}{\omega_{Bi}}
\newcommand{\omegap}{\omega_{\rm p}}
\newcommand{\omegavp}{\omega_{\rm vp}}
\newcommand{\Bq}{B_{\rm Q}}
\newcommand{\Rq}{R_{\rm Q}}
\newcommand{\mub}{\mu_B}
\newcommand{\thetab}{\theta_B}
\newcommand{\Thetab}{\Theta_B}
\newcommand{\polarb}{\beta}
\newcommand{\uel}{u_{\rm e}}
\newcommand{\uion}{u_{\rm i}}
\newcommand{\vel}{v}
\newcommand{\nudamp}{\nu_{\rm e}}
\newcommand{\nudampi}{\nu_{\rm i}}
\newcommand{\nudampre}{\nu_{\rm r,e}}
\newcommand{\nudampri}{\nu_{\rm r,i}}
\newcommand{\nudampce}{\nu_{\rm c,e}}
\newcommand{\nudampci}{\nu_{\rm c,i}}
\newcommand{\gamdampi}{\gamma_{\rm i}}

\newcommand{\gamdampri}{\gamma_{\rm r,i}}

\newcommand{\gamdampci}{\gamma_{\alpha,{\rm i}}}
\newcommand{\gamdampth}{\gamma_{\rm th}}

\newcommand{\opbrace}{\xi}
\newcommand{\lambdad}{\lambda_{s}}
\newcommand{\omegabo}{\omega_{Bs}}
\newcommand{\omegapo}{\omega_{{\rm p}s}}

\title{Atmospheres and Spectra of Strongly Magnetized Neutron Stars}
\author[W.C.G. Ho and D. Lai]{Wynn C. G. Ho and Dong Lai \\
Center for Radiophysics and Space Research, 
Department of Astronomy, Cornell University
Ithaca, NY 14853, USA \\
{\rm E-mail: wynnho@astro.cornell.edu, dong@astro.cornell.edu}}
\date{Accepted 2001 xxx,
      Received 2001 xxx;
      in original form 2001 xxx}

\pagerange{\pageref{firstpage}--\pageref{lastpage}}

\begin{document}

\maketitle

\label{firstpage}

\begin{abstract}
We construct atmosphere models for strongly magnetized neutron
stars with surface fields $B\sim 10^{12}-10^{15}$~G and
effective temperatures $T_{\rm eff}\sim 10^6-10^7$~K.
The atmospheres directly determine the characteristics of thermal
emission from isolated neutron stars, including radio pulsars,
soft gamma-ray repeaters, and anomalous X-ray pulsars.  In our models,
the atmosphere is composed of pure hydrogen or helium and is assumed
to be fully ionized.  The radiative opacities include
free-free absorption and scattering by both electrons and ions
computed for the two photon polarization modes in the magnetized
electron-ion plasma.  Since the radiation emerges from deep layers
in the atmosphere with $\rho\ga 10^2$~g/cm$^3$, plasma effects can
significantly modify the photon opacities by changing the
properties of the polarization modes.
In the case where the magnetic field and the surface normal are
parallel, we solve the full, angle-dependent, coupled radiative
transfer equations for both polarization modes.
We also construct atmosphere models for general field orientations
based on the diffusion approximation of the transport equations
and compare the results with models based on full radiative transport.
In general, the emergent thermal radiation exhibits significant
deviation from blackbody, with harder spectra at high energies.
The spectra also show a broad feature ($\Delta E/\Ebi\sim 1$)
around the ion cyclotron resonance
$\Ebi=0.63\,(Z/A)(B/10^{14}\mbox{ G})$~keV, where $Z$ and $A$
are the atomic charge and atomic mass of the ion, respectively;
this feature is particularly pronounced when $\Ebi\ga 3k\Teff$.
Detection of the resonance feature would provide a direct
measurement of the surface magnetic fields on magnetars.
\end{abstract}

\begin{keywords}
magnetic fields -- radiative transfer -- stars: atmospheres --
stars: magnetic fields -- stars: neutron -- X-rays: stars
\end{keywords}

\setcounter{equation}{0}
\section{Introduction} \label{sec:intro}

Neutron stars (NSs) are created in the core collapse and subsequent
supernova explosion of massive stars and begin their lives at
high temperatures, $T\ga 10^{11}$~K. As they cool over the next
$\sim 10^5-10^6$ years, they act as sources of soft X-rays
with surface temperatures $\ga 10^5$~K.
The cooling history of the NS depends on poorly constrained interior
physics such as the nuclear equation of state, superfluidity, and
magnetic field (see, e.g., Tsuruta~1998; Yakovlev \etal~2001, for a review)
Thus observations of surface emission from isolated NSs can
provide invaluable information on the physical properties
and evolution of NSs. In recent years, the spectra of several
radio pulsars (e.g., PSR~B1055-52, B0656+14, and Geminga) have
been observed to possess a thermal component that can be
attributed to the surface emission at temperatures in the range
$(2-10)\times 10^5$~K, while a few other pulsars
show thermal emission from hot spots at temperatures of a few times
$10^6$~K on the NS surface (e.g., for observations in the X-rays, see
Becker \& Tr\"{u}mper 1997; Becker 2000; in the extreme ultraviolet,
Edelstein \etal~1995; Korpela \& Bowyer~1998; in the optical band,
Pavlov \etal~1997; Caraveo \etal~2000). In addition, several
radio-quiet, isolated NSs, presumably accreting from the
interstellar medium (see Treves \etal~2000), have been
detected in the X-ray and optical bands
(e.g., Caraveo \etal~1996; Walter \etal~1996; Walter \& Matthews~1997;
Caraveo~1998; Mignani \etal~1998; Paerels \etal~2001).
Recent observations by the {\it Chandra} X-ray Observatory
have also revealed a number of compact central sources in supernova
remnants with spectra consistent with thermal emission
from isolated NSs\footnote{
see G. Pavlov 2000, http://online.itp.ucsb.edu/online/neustars\_c00/pavlov/}.
Most interestingly, in the last few years, thermal radiation has
been detected from a potentially new class of NSs, ``magnetars,''
which possess superstrong magnetic fields ($B\ga 10^{14}$~G)
(Duncan \& Thompson~1992; Thompson \& Duncan~1993, 1995, 1996;
see Thompson~2000 for a review).

Observations of soft gamma-ray repeaters (SGRs) and
anomalous X-ray pulsars (AXPs) strongly suggest that these
are NSs endowed with superstrong magnetic fields
(e.g., Vasisht \& Gotthelf~1997; Kouveliotou \etal~1998;
Kouveliotou \etal~1999;
see, e.g., Hurley~2000 for SGR review; Mereghetti~1999
for AXP review). According to the magnetar model, the X-ray
luminosity from AXPs and the quiescent X-ray emission from SGRs are
powered by the decay of a superstrong ($B\sim 10^{14}-10^{15}$~G)
magnetic field (Thompson \& Duncan~1996) and/or by the
residual thermal energy (Heyl \& Hernquist~1997a). Alternatively,
it has been proposed that the AXP emission may originate from
accretion onto a NS with $B\sim 10^{12}$~G, e.g., from a fossil
disk left over from the supernova explosion
(see Alpar~2000; Chatterjee \etal~2000, and references therein).
However, detailed timing studies of AXPs and SGRs
(e.g., Thompson \etal~2000; Kaspi \etal~2001) and deep optical observations
(e.g., Hulleman \etal~2000; Kaplan \etal~2001), as well as the many similarities
between AXPs and SGRs, clearly favor the magnetar model for both AXPs and
SGRs.  Nevertheless, it should be emphasized that, like radio pulsars, the
magnetic field strengths ($B\sim 10^{14}-10^{15}$~G) of magnetars are
indirectly inferred and are based on the measurement of the spin periods
($6-12$~s) and period derivatives and on the assumption that the spin-down
is due to the loss of angular momentum that is carried
away by magnetic dipole radiation or Alfv\'{e}n waves
(Thompson \& Blaes~1998; Thompson \etal~2000).
Thermal radiation has already been detected in four of the five
AXPs (1E~1048.1-5937, 1E~1845-0258, 4U0142+61, RXS~J1708-40)
and in SGR~1900+14; fits to the spectra with blackbody or with
crude atmosphere models indicate that the thermal X-rays can be
attributed to magnetar surface emission at temperatures of
$(3-7)\times 10^6$~K (see Mereghetti~1999; Perna \etal~2001).
Clearly, detailed observational and theoretical studies of
thermal emission can potentially reveal much about
the physical conditions and the true nature of magnetars. Of particular
importance are possible spectral features, such as the ion cyclotron
line [see eq.~(\ref{eq:ioncycen})]
studied in this paper (see also Zane \etal~2001), which, if detected,
can provide a direct measurement of the surface magnetic field.
In fact, it has been suggested that the lack of proton and
electron cyclotron features in {\it XMM-Newton} observations
of the isolated neutron star RX~J0720.4-3125 may lead to
constraints on its surface magnetic field (Paerels \etal~2001).

Thermal radiation from a magnetized NS is mediated by the thin
atmospheric layer (with scale height $\sim 0.1-10$~cm and
density $\sim 0.1-100$~g/cm$^3$) that covers the stellar surface.
The physical properties of the atmosphere, such as the
chemical composition, equation of state, and especially the radiative
opacities, directly determine the characteristics of the thermal
emission.  While the surface composition of the NS is unknown,
a great simplification arises due to the efficient gravitational
separation of light and heavy elements
(with timescales on the order of seconds; see Alcock \& Illarionov~1980
for a discussion of gravitational separation in white dwarfs).
A pure H atmosphere is expected even if a small
amount of fallback/accretion occurs after NS formation\footnote{
As far as the radiation spectrum is concerned, the photons decouple
from the matter at Thomson depth $\tau\la 10^3$ (see Fig.~\ref{fig:taunu});
the total mass above this depth is less than $10^{16}$~g.}.
A pure He atmosphere results if H is completely burnt up,
and a pure Fe atmosphere may be possible if no fallback/accretion occurs.

Steady progress has been made over the years in modeling NS
atmospheres.  The first models of zero-field NS atmospheres
were constructed by Romani~(1987). Later works used improved
opacity and equation of state data from the OPAL project for
pure hydrogen, helium, and iron compositions
(Rajagopal \& Romani~1996; Zavlin \etal~1996).  These works showed
that the radiation spectra from light element (H or He),
low-field ($B\la 10^8$~G) atmospheres deviate significantly
from blackbody spectra. Magnetic NS atmosphere modeling
was first attempted by Miller~(1992), who unfortunately adopted an
incorrect polarization-averaging procedure for the radiative transport
and included only bound-free opacities.  So far the most comprehensive
studies of magnetic NS atmospheres have focused
on hydrogen and moderate field strengths of $B\sim 10^{12}-10^{13}$~G
(Shibanov \etal~1992; Pavlov \etal~1994; see Pavlov \etal~1995
for a review and Zane \etal~2000
for full, angle-dependent atmosphere models with accretion).
These models correctly take into account the transport of
different photon modes (in the diffusion approximation)
through a mostly ionized medium.  The opacities adopted in the
models include free-free transitions and electron scattering,
while bound-free opacities are treated in a highly approximate
manner and bound-bound transitions are completely ignored.
Because the strong magnetic field significantly increases
the binding energies of atoms, molecules, and other bound states
(see Lai~2001 for a recent review), these bound states have
appreciable abundances in the atmosphere at low temperatures
($T\la 10^6$~K) (see Lai \& Salpeter~1997; Potekhin \etal~1999).
Therefore the models of Pavlov et~al.\ are expected
to be valid only for relatively high temperatures
($T\ga {\rm a~few}~\times 10^6$~K), where hydrogen is almost
completely ionized.
The presence of a magnetic field gives rise to anisotropic and polarized
emission from the atmosphere (see Pavlov \& Zavlin~2000).
The resulting magnetic spectra are softer at X-ray energies
than the low-field spectra but are still harder than the blackbody
spectra.  Models of magnetic iron atmospheres (with $B\sim 10^{12}$~G)
were studied by Rajagopal \etal~(1997). Because of the complexity
in the atomic physics and radiative transport, these Fe
models are necessarily crude. Despite a number of shortcomings,
these H and Fe atmosphere models
have played a valuable role in assessing the observed spectra
of several radio pulsars and radio-quiet NSs (e.g., it is now
well recognized that fitting the thermal spectra by a blackbody
in the X-rays tends to overestimate the effective temperature)
and some useful constraints on NS properties have been
obtained (e.g., Meyer \etal~1994; Pavlov \etal~1996; Zavlin \etal~1998).

This paper is the first in a series where we systematically
investigate the atmosphere and spectra of strongly magnetized NSs.
Here we study the H and He atmospheres with relatively high effective
temperatures ($T_{\rm eff}\ga {\rm a~few}\times 10^6$~K) so that bound atoms
or molecules may be neglected. We focus on the superstrong field regime
($B\ga 10^{14}$~G) and compare with results from models possessing
weaker magnetic fields.
We construct self-consistent atmosphere models in radiative equilibrium
and solve the full, angle-dependent, coupled transport equations for both
polarization modes. For comparisons, we also study
models based on the diffusion approximation of the transport equations.
Of particular interest is the effect of the ion cyclotron resonance,
which occurs at energy
\be
\Ebi=\hbar\omegabi = \hbar\frac{ZeB}{A\mpr c}
= 0.63\lp\frac{Z}{A}\rp\lp\frac{B}{10^{14}{\rm G}}\rp\mbox{keV},
\label{eq:ioncycen}
\ee
where $Z$ and $A$ are the charge number and mass number of
the ion, respectively.
As we show in this paper, the resonance gives rise to
a broad ($\Delta E/E\sim 1$) spectral feature around $E_{Bi}$,
and this feature is particularly pronounced when
$E_{Bi}\ga 3\,kT_{\rm eff}$. Obviously, the ion cyclotron
feature can potentially provide important diagnostics for
the physical properties of magnetars.

While in the final stages of writing our paper for publication, we
became aware of two recent papers that also explore H atmospheres
in the superstrong field regime. \"{O}zel~(2001) included the effect
of vacuum polarization (see Section~\ref{sec:discussion}) but
neglected the ion effect on both
opacities and polarization modes, while Zane \etal~(2001) studied
the proton cyclotron feature and adopted an approximate description
of the polarization modes in the electron-proton plasma
with vacuum polarization (see also Section~\ref{sec:discussion}).

In Section~\ref{sec:atmmodel}, we review the basic physics
ingredients of our atmosphere models, including the radiative transfer
equations, the equation of state, and the opacities.
The numerical methods for both full transport and diffusion models
and tests of the solution are discussed in Section~\ref{sec:numcomp}.
In Section~\ref{sec:results}, we present our numerical
atmosphere models and spectra for different compositions (H or He),
effective temperatures, and magnetic field strengths and
geometries. Finally, Section~\ref{sec:discussion} contains a
discussion of the uncertainties in our models and possible
future works.

\setcounter{equation}{0}
\section{Atmosphere Model: Basic Equations and Physics Inputs}
\label{sec:atmmodel}

We consider an isolated NS with a plane-parallel
atmosphere.  This is justified since the atmospheric scale
height $H \la 10$~cm is much less than the NS radius
$R\approx 10$~km.  The atmosphere is composed of pure
hydrogen or helium (see Section~\ref{sec:intro}).
The temperature $T$ is taken to be
$\ga 10^6$~K, and the atmosphere is assumed to be fully ionized
(see Section~\ref{sec:discussion} for discussion on ionization
equilibrium).
A uniform magnetic field $\vecB$ permeates the atmosphere, and
the angle between the direction of the magnetic field $\hat{B}$ and
the surface normal $\hat{z}$ is $\Thetab$.

\subsection{Radiative Transfer Equation} \label{sec:rte}

In a magnetized plasma, there are two normal modes of propagation
for electromagnetic waves.  These are the extraordinary mode
(X-mode, $j = 1$), which is mostly polarized perpendicular to the
$\veck$-$\vecB$ plane, and the ordinary mode (O-mode, $j = 2$), which
is mostly polarized parallel to the $\veck$-$\vecB$ plane,
where $\veck$ is the unit vector along the wave propagation
direction (e.g., M\'{e}sz\'{a}ros~1992; see Section~\ref{sec:opacity0}).
When the normal modes are approximately orthogonal and their
relative phase shift over a mean free path is large (i.e., large
Faraday depolarization), the radiative transfer of the four
Stokes parameters reduces to that of the two modes
(see Gnedin \& Pavlov~1974).
The radiative transfer equation (RTE) for the specific
intensity $\Inu^j$ of mode $j$ is
\be
\veck\cdot\nabla\Inu^j(\veck)=\rho\kabsj(\veck)\frac{\Bnu}{2}
-\rho\ktotj(\veck)\Inu^j(\veck)+\rho\sum_{i=1}^2\int\!d\veck'
 \frac{d\ksc(\veck' i\rightarrow\veck j)}{d\Omega}\Inu^i(\veck'),
\label{eq:rte0}
\ee
where $\Bnu=\Bnu(T)$ is the blackbody intensity, $\kabsj(\veck)$
(in cm$^2$/g) is the absorption opacity (corrected for stimulated
emission) for mode $j$ propagating
along $\veck$,
$d\ksc(\veck' i\rightarrow\veck j)/d\Omega$
is the differential opacity for scattering from mode $i$
in direction $\veck'$ to mode $j$ in direction $\veck$,
$\ktotj(\veck)=\kabsj(\veck)+\kscj(\veck)$ is the total
opacity, and
\ba
\kscj(\veck) & = & \sum_{i=1}^2\kscji(\veck) \\
\kscji(\veck) & = & \int\!d\veck'
 \frac{d\ksc(\veck j\rightarrow\veck' i)}{d\Omega'}.
\ea

Defining the Thomson depth $\tau$ by $d\tau = -\rho\keso dz$,
where $\keso = \Ye\sigth/\mpr = 0.4\,\Ye$~cm$^2$/g is the Thomson
electron scattering opacity in the absence of a magnetic field
and $\Ye=Z/A$ is the electron fraction,
equation~(\ref{eq:rte0}) becomes
\be
\mu\frac{\partial\Inu^j(\veck)}{\partial\tau}=\frac{\ktotj(\veck)}{\keso}
 \Inu^j(\veck)-\frac{\kabsj(\veck)}{\keso}\frac{\Bnu}{2}-\frac{1}{\keso}
 \sum_{i=1}^2\int\!d\veck'
 \frac{d\ksc(\veck' i\rightarrow\veck j)}{d\Omega}\Inu^i(\veck'),
\label{eq:rte2}
\ee
where $\mu\equiv\veck\cdot\hat{z}=\cos\theta$.
In general, at a given frequency $\nu$ and depth $\tau$,
the intensity $\Inu^j(\veck)$ depends on $\mu$ and the
azimuthal angle $\phi$ when $\vecB$ is not along the
surface normal, i.e., when $\Thetab\ne 0$.
In solving the RTE, it is useful to introduce the two
quantities
\ba
\meani(\veck) & = & \frac{1}{2}[\Inu^j(\veck)+\Inu^j(-\veck)] \\
\meanf(\veck) & = & \frac{1}{2}[\Inu^j(\veck)-\Inu^j(-\veck)],
\ea which are
evaluated only for $\mu\ge 0$ and are related to the mean
specific intensity and energy flux
(see Mihalas~1978 for the nonmagnetic case).
As a result, the RTE becomes
\ba
\meanf(\veck) & = & \mu\frac{\partial\meani(\veck)}{\partial\tau_\nu}
 \label{eq:rte1b} \\
\mu\frac{\partial\meanf(\veck)}{\partial\tau_\nu} & = &
 \mu^2\frac{\partial^2\meani(\veck)}{\partial\tau_\nu^2} = \meani(\veck)
 - \frac{\kabsj(\veck)}{\ktotj(\veck)}\frac{\Bnu}{2}
 - \frac{2}{\ktotj(\veck)} \sum_{i=1}^2\int_+\!d\veck'
 \frac{d\ksc(\veck' i\rightarrow\veck j)}{d\Omega} i_\nu^i(\veck'),
 \label{eq:rte1}
\ea
where $d\tau_\nu = -\rho\ktotj(\veck)dz$ is the total optical
depth and $\int_+$ implies that the domain of integration
is restricted to $\veck'\cdot\hat{z}=\mu'\ge 0$.
In deriving these equations, we have used the reflection
symmetry of the opacities, i.e.,
$\ktotj(\veck)=\ktotj(-\veck)$ and
$d\ksc(\veck' i\rightarrow\veck j)/d\Omega
=d\ksc(\veck' i\rightarrow-\veck j)/d\Omega
=d\ksc(-\veck' i\rightarrow-\veck j)/d\Omega$.
The boundary conditions for the RTE are $\Inu^j(\veck)=0$ for
$\mu < 0$ at $\tau_\nu=0$ and $\Inu^j(\veck)\rightarrow\Bnu/2$
at $\tau_\nu\rightarrow\infty$.  In terms of $\meani$, these are
\ba
\mu\frac{\partial\meani}{\partial\tau_\nu} & = & \meani
 \qquad \,\, \mbox{at } \tau_\nu=0 \label{eq:bc1} \\
\mu\frac{\partial\meani}{\partial\tau_\nu} + \meani & = &
\frac{1}{2}\lp\mu\frac{\partial\Bnu}{\partial\tau_\nu} + \Bnu\rp
 \qquad \,\, \mbox{at } \tau_\nu\rightarrow\infty. \label{eq:bc2}
\ea

To obtain the constraint imposed by radiative equilibrium, we
integrate the RTE~(\ref{eq:rte0}) over the solid angle, which
yields the zeroth-order moment of the transfer equation,
\be
\nabla\cdot\vFnu^j(\veck)=\rho\int d\veck\,\kabsj(\veck)\frac{\Bnu}{2}
-\rho\int d\veck\,\ktotj(\veck)\Inu^j(\veck)+\rho\sum_{i=1}^2
 \int d\veck\int d\veck'
 \frac{d\ksc(\veck' i\rightarrow\veck j)}{d\Omega}\Inu^i(\veck'),
\label{eq:radeq0}
\ee
where the specific energy flux of mode $j$ is
\be
\vFnu^j\equiv\int d\veck\,\veck\Inu^j(\veck)
 = 2\int_+ d\veck\,\veck\meanf(\veck).
\ee
In radiative equilibrium,
the total flux, $\vecF=\sum_{j=1}^2\int d\nu\,\vFnu^j$,
satisfies $\nabla\cdot\vecF=0$, so that
\be
\nabla\cdot\vecF=2\rho\sum_{j=1}^2\int d\nu\int_+ d\veck\,\kabsj(\veck)
 \lb \frac{\Bnu}{2} - \meani(\veck)\rb = 0. \label{eq:radeq}
\ee
Also, for an atmosphere with an effective temperature $\Teff$,
the total flux is constant and satisfies
\be
F_z = \sum_{j=1}^2\int d\nu\int d\veck\,\mu\Inu^j(\veck)
 = 2\sum_{j=1}^2\int d\nu\int_+ d\veck\,\mu\meanf(\veck) = \sigsb\Teff^4,
\label{eq:fluxconst}
\ee
where $\sigsb$ is the Stefan-Boltzmann constant.

\subsection{Special Case: Magnetic Field Parallel to Surface Normal}
\label{sec:thetab0}

In the special case when $\vecB$ is parallel to the surface
normal $\hat{z}$, the intensity $\Inu^j(\veck)$ is independent
of $\phi$, so that $\Inu^j(\veck)=\Inu^j(\mu)$.  In the
atmosphere models based on the full RTE presented in this
paper, we consider only this special case.  It is clear that
the RTE~(\ref{eq:rte1}) can be written in the familiar form
\be
\mu^2\frac{\partial\meani}{\partial\tau_\nu}=\meani-\Snu^j,
 \label{eq:rtestandard}
\ee
where
\be
\Snu^j(\mu) = \frac{\kabsj(\mu)}{\ktotj(\mu)} \frac{\Bnu}{2}
 + \frac{4\pi}{\ktotj(\mu)} \sum_{i=1}^2 \int_0^1 d\mu'
 \frac{d\ksc(\mu' i\rightarrow\mu j)}{d\Omega} i_\nu^i(\mu')
\label{eq:source}
\ee
is the source function.

We note that, in using equation~(\ref{eq:source}) to calculate
the source function, the integral over the initial photon
direction $\mu'$ must be carried out at every frequency $\nu$,
depth $\tau$, and final photon direction $\mu$ and for each global
iteration (temperature correction) as described in
Section~\ref{sec:numcomp}.  We can reduce the computation time
by assuming that the differential scattering cross-section is 
approximately independent of the initial photon direction.
Equation~(\ref{eq:source}) then yields the approximate source function
\be
\Snu^j(\mu) \approx \frac{\kabsj(\mu)}{\ktotj(\mu)} \frac{\Bnu}{2}
 +
\sum_{i=1}^2 \frac{\kscji(\mu)}{\ktotj(\mu)}\frac{c\unu^i}{4\pi},
\label{eq:sourceapprox}
\ee
where the specific energy density is
\be
\unu^j=\frac{1}{c}\int d\veck\,\Inu^j(\veck).
\ee
We expect that this approximation introduces only a small error to
the emergent spectra, and a comparison of the final results using
the source functions given by equations~(\ref{eq:source})
and (\ref{eq:sourceapprox}) indeed show the differences are
small, though this has not been studied extensively
(see also Shibanov \& Zavlin~1995).
The numerical results presented in Section~\ref{sec:results}
make use of the approximate source function given by
equation~(\ref{eq:sourceapprox}).

\subsection{Diffusion Approximation} \label{sec:rted}

In this paper, we shall also present atmosphere models for general
$\Thetab$ based on the diffusion approximation of the RTE.
To obtain the diffusion RTE, we
multiply both sides of the RTE~(\ref{eq:rte0}) by
$-\veck/\rho\ktotj(\veck)$ and integrate over the solid
angle to find the specific flux,
\be
\vFnu^j=-\int d\veck\frac{1}{\rho\ktotj(\veck)}
 \veck(\veck\cdot\nabla)\Inu^j(\veck), \label{eq:spflux}
\ee
where we have used the symmetry property of the opacities,
so that $\int d\veck\,[\veck/\ktotj(\veck)]
 [d\ksc(\veck' i\rightarrow\veck j)/d\Omega]=0$ and \\
$\int d\veck\,\veck[\kabsj(\veck)/\ktotj(\veck)]=0$.
In the diffusion limit, we approximate the specific intensity by
$\Inu^j(\veck)\approx(c/4\pi)[\unuj+(3/c)\veck\cdot\vFnu^j]$.
\label{eq:spintensd}
Substituting into equation~(\ref{eq:spflux}) and using the fact
that $\nabla\unuj=(\partial\unuj/\partial z)\hat{z}$ and that
the opacity only depends on $\mub=\veck\cdot\hat{B}=\cos\thetab$,
we find
\be
\vFnu^j\approx-c\frac{\partial\unuj}{\partial z}
 \lp\mfpj^\parallel\cos\Thetab\hat{B}+\mfpj^\perp\sin\Thetab\hat{B}_\perp\rp,
\label{eq:spfluxd}
\ee
where $\hat{B}_\perp$ is the unit vector perpendicular to
$\hat{B}$ and lies in the plane of $\hat{B}$ and $\hat{z}$
(recall that $\Thetab$ is the angle between $\hat{B}$ and $\hat{z}$)
and $\hat{B}_\perp\sin\Thetab=\hat{z}-(\hat{z}\cdot\hat{B})\hat{B}$.
In equation~(\ref{eq:spfluxd}), $\mfpj^\parallel$ and $\mfpj^\perp$
are the angle-averaged mean free paths parallel and perpendicular
to the magnetic field, respectively, and they are given by
\ba
\mfpj^\parallel & = & \int_0^1 d\mub\frac{\mub^2}{\rho\ktotj(\mub)} \\
\mfpj^\perp & = & \frac{1}{2}\int_0^1 d\mub\frac{(1-\mub^2)}{\rho\ktotj(\mub)}.
\ea
The specific flux perpendicular to the stellar surface is
$F_{\nu,z}^j\approx-c\mfpj(\partial\unuj/\partial z)$,
where the averaged mean free path along $z$ is
\be
\mfpj = \mfpj^\parallel \cos^2\!\Thetab + \mfpj^\perp \sin^2\!\Thetab.
 \label{eq:defnmfpj}
\ee
Note that, as a consequence of equation~(\ref{eq:spfluxd}),
a magnetic field can induce horizontal radiative flux; this implies
that a horizontal temperature gradient may develop in the
atmosphere.  To properly address this issue, one needs to
study the three-dimensional structure of the atmosphere;
this is beyond the scope of this paper (see Section~\ref{sec:discussion}).
Substituting the approximate expressions of the specific intensity and flux
into the zeroth-order moment equation~(\ref{eq:radeq0})
and defining the Thomson depth $d\tau=-\rho\keso\,dz$ and the
zero-field Thomson mean free path $\mfpo\equiv 1/(\rho\keso)$, we obtain
\be
\frac{\partial}{\partial\tau}\lp\frac{\mfpj}{\mfpo}
 \frac{\partial\unuj}{\partial\tau}\rp
 = \frac{\Kabsj}{\keso}\lp\unuj-\frac{\unup}{2}\rp 
 + \frac{\Kexc}{\keso}\lp\unuj-\unu^{3-j}\rp, \label{eq:rte1d}
\ee
where $\unup=(4\pi/c)\Bnu(T)$ is the blackbody specific energy density and
\ba
\Kabsj & = & \frac{1}{4\pi}\int d\veck\,\kabsj(\veck) \label{eq:kabsjavg} \\
\Ktotj & = & \frac{1}{4\pi}\int d\veck\,\ktotj(\veck) \label{eq:ktotjavg} \\
\Kscji & = & \frac{1}{4\pi}\int d\veck\,\kscji(\veck). \label{eq:kscjiavg}
\ea
This equation has been previously derived by
Kaminker \etal~(1982, 1983).

The boundary conditions for equation~(\ref{eq:rte1d}) are
$\hat{z}\cdot\vFnu^j=c\unuj/2$ at $\tau=0$ and
$\unuj\rightarrow\unup/2$ at $\tau\rightarrow\infty$, i.e.,
\ba
\frac{\mfpj}{\mfpo}\frac{\partial\unuj}{d\tau} & = &
 \frac{\unuj}{2} \qquad \,\, \mbox{at } \tau=0 \label{eq:bc1d} \\
\frac{\partial\unuj}{\partial\tau} + \unuj & = &
 \frac{1}{2}\lp\frac{\partial\unup}{\partial\tau} + \unup\rp
 \qquad \mbox{at } \tau \rightarrow \infty. \label{eq:bc2d}
\ea
Finally, in the diffusion approximation, the radiative
equilibrium condition [see eq.~(\ref{eq:radeq})] simplifies to
\be
\frac{dF_z}{d\tau} = c\sum_{j=1}^2\int d\nu\,\frac{\Kabsj}{\keso}
 \lp\unuj-\frac{\unup}{2}\rp = 0. \label{eq:radeqd}
\ee

\subsection{Hydrostatic Equilibrium and Equation of State} \label{sec:eos}

In modeling the atmosphere, we must determine the
density $\rho(\tau)$ and temperature $T(\tau)$ along with
the radiation intensity.
From hydrostatic equilibrium, the pressure $P$ at Thomson
depth $\tau$ is given by
\be
P(\tau) = \frac{g}{\keso}\,\tau, \label{eq:ptau}
\ee
where $g = (GM/R^2)(1-2GM/Rc^2)^{-1/2} = 2.4\times 10^{14}$~cm/s$^2$
is the gravitational acceleration at the surface of a
$M$=1.4~M$_\odot$, $R$=10~km neutron star.
We assume the atmosphere is fully ionized.
The equation of state can be written as
\be
P = P_{\rm e} + P_{\rm ion} + P_{\rm Coul}, \label{eq:prestot}
\ee
where $P_{\rm e}$ is the pressure due to electrons, which may
or may not be degenerate, $P_{\rm ion}$ is the pressure due
to ions, which still behave as a classical ideal gas under the
conditions that exist in the upper layers of the NS,
and $P_{\rm Coul}$ is the Coulomb correction to the pressure.
In strong magnetic fields, electrons are quantized into Landau
levels, and the electron pressure is given by (e.g., Lai~2001)
\be
P_{\rm e} = \frac{\me c^2}{2^{1/2}\pi^2\lambdae^{3}}
 \frac{B}{\Bq} \lp\frac{kT}{\me c^2}\rp^{3/2}
 \sum_{\nu=0}^\infty g_\nu F_{1/2}\lp\frac{\mu_{\rm e}-\nu\hbar\omegab}{kT}\rp,
\label{eq:presel}
\ee
where $\lambdae=\hbar/\me c$ is the electron Compton wavelength,
$\Bq=\me^2c^3/(e\hbar)=4.414\times 10^{13}$~G, $g_\nu$ is the
spin degeneracy of the Landau level ($g_0=1$ and $g_\nu=2$ for
$\nu\ge 1$), $\mu_{\rm e}$ is the
electron chemical potential, and $F_n(y)$ is the Fermi-Dirac integral
\be
F_n(y) \equiv \int_0^\infty\frac{x^n}{e^{x-y}+1}dx.
\ee
When the conditions
\ba
T & \ll & T_B = \frac{\hbar\omegab}{k} = 1.34\times 10^8\,B_{12}\mbox{ K} \\
\rho & \le & \rho_B = \frac{\mpr}{\sqrt{2}\pi^2\Ye\Rq^3}
 = 7.05\times 10^3\,\Ye^{-1} B_{12}^{3/2}\mbox{ g/cm$^3$},
\ea
where $\Rq=(\hbar c/eB)^{1/2}$ and $B_{12}=B/$(10$^{12}$~G), are satisfied,
as in the case of the atmospheres considered
in this paper, only the ground Landau level ($\nu = 0$) is occupied.
Thus, in equation~(\ref{eq:presel}), only the first term in the
sum need be considered.  The electron number density $\nel$ is given by
\be
\nel = \frac{1}{2^{3/2}\pi^2\lambdae^{3}}\frac{B}{\Bq}
\lp\frac{kT}{\me c^2}\rp^{1/2} F_{-1/2}\lp\frac{\mu_{\rm e}}{kT}\rp,
 \label{eq:ndens}
\ee
where we have kept only the $\nu=0$ term.  For $\rho\le\rho_B$,
the electron Fermi temperature is
\be
T_{\rm F} = E_{\rm F}/k = 2.70\,B_{12}^{-2}\lp\Ye\rho_1\rp^2\mbox{ K},
 \qquad \mbox{(for $\rho\le\rho_B$)} \label{eq:tempfermi}
\ee
where $\rho_1=\rho/$(1~g cm$^{-3}$).  Because electrons are confined by
the magnetic fields to smaller volumes, electron degeneracy occurs
at higher densities than in the absence of the magnetic field.
Note that for $T\gg T_{\rm F}$, equation~(\ref{eq:presel}) reduces
to $P_{\rm e}\approx\nel kT$ even when the field is strongly
quantizing ($\rho\le\rho_B$).
The ion pressure, on the other hand, is well approximated by
\be
P_{\rm ion} = \frac{\rho kT}{A\mpr}. \label{eq:presideal}
\ee
The Coulomb correction is given by
(e.g., Shapiro \& Teukolsky~1983)
\be
P_{\rm Coul} = - \frac{3}{10}\lp\frac{4\pi}{3}\rp^{1/3}Z^{2/3}e^2\nel^{4/3}.
\ee

The equation of state, including the effects of degeneracy,
is determined by substituting the different components of the
pressure into equation~(\ref{eq:prestot}) to obtain a relation
between a given total pressure $P(\tau)$ and temperature $T(\tau)$.
The root of this equation yields the degeneracy parameter
$\eta(\tau)=\mu_{\rm e}/(kT)$
which is then substituted into equation~(\ref{eq:ndens})
to obtain the electron number density $\nel$ and the total density
$\rho = (A\mpr/Z)\nel$.  We use the rational function approximation
given by Antia~(1993) to evaluate the Fermi-Dirac integrals.
For all the nonmagnetic and
magnetic atmosphere models considered here, the difference between
the densities calculated from the ideal gas equation of state
and the equation of state including electron degeneracy is only
a few percent in the innermost layers, and the effect on the surface
spectrum is negligible (see Section~\ref{sec:results}).

\subsection{Opacities} \label{sec:opacity0}

In a magnetized electron-ion plasma, the scattering and free-free
absorption opacities depend on the direction of propagation
$\veck$ (more precisely, on the angle $\thetab$ between $\veck$ and
$\hat{B}$) and the normal modes ($j = 1,2$) of the electromagnetic
waves.  In rotating coordinates, with the $z$-axis along $\hat{B}$,
the components of the polarization vector $\vece^j$ are given
by (see Appendix~\ref{sec:polarvec})
\ba
|e_\pm^j|^2 & = & \left|\frac{1}{\sqrt{2}}\lp e_x^j\pm ie_y^j\rp\right|^2
 = \frac{1}{2(1+K_j^2+K_{z,j}^2)}
 \lb 1\pm\lp K_j\cos\thetab+K_{z,j}\sin\thetab\rp\rb^2 \label{eq:polarvecpm} \\
|e_z^j|^2 & = & \frac{1}{1+K_j^2+K_{z,j}^2}
 \lp K_j\sin\thetab-K_{z,j}\cos\thetab\rp^2
 \label{eq:polarvecz},
\ea
where
\ba
K_j & = & \polarb\lb 1+(-1)^j \lp 1+\frac{1}{\polarb^2}\rp^{1/2}\rb
 \label{eq:polarkj} \\
K_{z,j} & = & \frac{\uel\vel\lp 1-\uion-M^{-1}\rp
 \sin\thetab\cos\thetab K_j - \uel^{1/2}\vel\sin\thetab}{\lp 1-\uel\rp
 \lp 1-\uion\rp - \vel\lb\lp 1-\uion\rp\lp 1-\uel\cos^2\thetab\rp
 - M\uion\sin^2\thetab\rb} \label{eq:polarkz} \\
\polarb & = & \frac{\uel^{1/2}}{2(1-\vel)}\frac{\sin^2\!\thetab}{\cos\thetab}
  \lp 1-\uion-\frac{ 1+\vel}{M}\rp,
\label{eq:polarb}
\ea
and $M\equiv A\mpr/(Z\me)$.
In equations~(\ref{eq:polarkj})-(\ref{eq:polarb}), we have defined
\be
\uel = \frac{\omegab^2}{\omega^2}, \qquad
\uion = \frac{\omegabi^2}{\omega^2}, \qquad \vel = \frac{\omegap^2}{\omega^2};
\ee
the electron cyclotron frequency
$\omegab$, the ion cyclotron frequency $\omegabi$, and the electron
plasma frequency $\omegap$ are given by
\ba
\hbar\omegab & = & \hbar\frac{eB}{\me c}
 = 11.58\,B_{12}\mbox{ keV}  \\
\hbar\omegabi & = & \hbar\frac{ZeB}{A\mpr c}
 = 6.305\,B_{12}\lp\frac{Z}{A}\rp\mbox{ eV} \\
\hbar\omegap & = & \hbar\lp\frac{4\pi e^2\nel}{\me}\rp^{1/2}
 = 28.71\,\lp\frac{Z}{A}\rp^{1/2}\rho_1^{1/2}\mbox{ eV},
\ea
respectively.
Note that in the coordinate frame where the $z$-axis is along
$\veck$ and the $x$-axis is in the $\veck$-$\vecB$ plane, the
polarization vector can be written as (see Appendix~\ref{sec:polarvec})
\be
\vece^j = \frac{1}{\sqrt{1+K_j^2+K_{z,j}^2}}(iK_j,1,iK_{z,j}).
 \label{eq:polarveck}
\ee
Thus the modes are elliptically polarized with $K_j$ being the ratio
of the axes of the ellipse.  It is easy to show that
$K_1 = -K_2^{-1}$ and $|K_1| < 1$; this implies that the
extraordinary mode ($j=1$) is mostly perpendicular to the
$\veck$-$\vecB$ plane, while the ordinary mode ($j=2$) lies mainly
in the $\veck$-$\vecB$ plane.
For $\omega \ll \omegab$, the modes are nearly linearly polarized
at most angles except close to parallel propagation $\thetab = 0$ and $\pi$,
where the polarization is circular since $K_j=(-1)^j$.
In this limit, $|e_\pm^j|^2$ and $|e_z^2|^2\sim 1$, and
$|e_z^1|^2\sim\polarb^{-2}$ [see Section~\ref{sec:opacity},
specifically eqs.~(\ref{eq:polarvec1approx}) and
(\ref{eq:polarvec2approx})].

It is evident from equation~(\ref{eq:polarkz}) that
the component of $\vece^j$ along $\veck$ is of order
$\vel\propto\rho/\omega^2$, and thus, at sufficiently low
densities, $K_{z,j}$ can be neglected so that the modes
are transverse.  Previous studies adopted this ``transverse-mode''
approximation when treating the radiative opacities in a
magnetized medium (e.g., Pavlov \etal~1995).  As we
shall see in Section~\ref{sec:results}, since the
radiation can directly emerge from deep layers where $\vel\ga 1$,
the transverse-mode approximation cannot always be justified.

In a magnetized medium, the electron scattering opacity from
mode $j$ into mode $i$ is given by (Ventura~1979; Ventura, Nagel, \&
M\'{e}sz\'{a}ros~1979)
\be
\kesji = \frac{\nel\sigth}{\rho}\sum_{\alpha=-1}^1
 \frac{\omega^2}{(\omega+\alpha\omegab)^2+\nudamp^2}|e_\alpha^j|^2A_\alpha^i,
 \label{eq:kesji}
\ee
where $A_\alpha^i$ is the angle integral given by
\be
A_\alpha^i = \frac{3}{8\pi}\int d\veck\,|e_\alpha^i|^2.
\ee
The electron scattering opacity from mode $j$ is
\be
\kesj = \frac{\nel\sigth}{\rho}\sum_{\alpha=-1}^1
 \frac{\omega^2}{(\omega+\alpha\omegab)^2+\nudamp^2}|e_\alpha^j|^2 A_\alpha,
 \label{eq:kesj}
\ee
where $A_\alpha=\sum_{i=1}^2 A_\alpha^i$.  In the transverse-mode
approximation ($K_{z,j}=0$), the polarization vector $\vece^j$
satisfies the completeness relation
$\sum_{j=1}^2 |e_\pm^j|^2 = \lp 1+\cos^2\!\thetab\rp/2$
and $\sum_{j=1}^2 |e_z^j|^2 = \sin^2\!\thetab$, and thus $A_\alpha=1$.
Note that a resonance develops near the electron cyclotron
frequency for the circularly polarized mode with its electric
field rotating in the same direction as the electron gyration;
however, this resonance appears only in the extraordinary mode
since one can easily show that $|e_-^2|^2\approx 0$ at $\omega=\omegab$.
For the photon energies of interest in this paper, i.e.,
$\omega\ll\omegab$, a suppression factor $(\omega/\omegab)^2$
in the opacity
results from the strong confinement of electrons perpendicular
to the magnetic field.  Similar features appear in the electron
free-free absorption opacity
(e.g., Virtamo \& Jauho~1975; Pavlov \& Panov~1976; Nagel \& Ventura~1983)
\be
\kffj = \frac{\alpha_0}{\rho}\sum_{\alpha=-1}^1
 \frac{\omega^2}{(\omega+\alpha\omegab)^2+\nudamp^2}|e_\alpha^j|^2\gff_\alpha,
 \label{eq:keffj}
\ee
where
\be
\alpha_0 = 4\pi^2 Z^2\alphaf^3\frac{\hbar^2 c^2}{\me^2}
 \lp\frac{2\me}{\pi kT}\rp^{1/2} \frac{\nel\nion}{\omega^3}
 \lp 1-e^{-\hbar\omega/kT}\rp = \alpha^{\rm ff}_0 \frac{3\sqrt{3}}{4\pi}
 \frac{1}{\gff}, \label{eq:kffmagcoef}
\ee
$\alphaf=e^2/(\hbar c)$ is the fine structure constant, and
$\alpha^{\rm ff}_0$ and $\gff$ are the free-free
absorption coefficient and velocity-averaged free-free Gaunt
factor, respectively, in the absence of a magnetic field.
In equation~(\ref{eq:keffj}), $\gff_{\pm 1}=\gff_\perp$ and
$\gff_0=\gff_\parallel$ are the velocity-averaged free-free
Gaunt factors in a magnetic field, which we evaluate using
the expressions given in M\'{e}sz\'{a}ros~(1992)
(see also Nagel~1980; note these expressions for the Gaunt factors
neglect transitions to the excited electron Landau levels;
see Pavlov \& Panov~1976 for the complete expressions).

In an ionized medium, the ions also contribute to the total
scattering and absorption opacities.  Analogous to
equations~(\ref{eq:kesji})-(\ref{eq:keffj}), we have
\ba
\kisji & = & \lp\frac{Z^2\me}{A\mpr}\rp^2 \frac{\nion\sigth}{\rho}
 \sum_{\alpha=-1}^1\frac{\omega^2}{(\omega-\alpha\omegabi)^2+\nudampi^2}
 |e_\alpha^j|^2A_\alpha^i \label{eq:kisji} \\
\kisj & = & \lp\frac{Z^2\me}{A\mpr}\rp^2\frac{\nion\sigth}{\rho}
 \sum_{\alpha=-1}^1\frac{\omega^2}{(\omega-\alpha\omegabi)^2+\nudampi^2}
 |e_\alpha^j|^2 A_\alpha \label{eq:kisj} \\
\kiffj & = & \frac{1}{Z^3}\lp\frac{Z^2\me}{A\mpr}\rp^2\frac{\alpha_0}{\rho}
 \sum_{\alpha=-1}^1\frac{\omega^2}{(\omega-\alpha\omegabi)^2+\nudampi^2}
 |e_\alpha^j|^2\gff_\alpha. \label{eq:kiffj}
\ea
Note that the ion cyclotron resonance occurs for $\alpha=+1$, i.e.,
when the electric field of the mode rotates in the same direction
as the ion gyration.
The total scattering and absorption opacities in a fully ionized
medium is then the sum of the electron and ion components, namely
$\kscj = \kesj + \kisj$ and $\kabsj = \kffj + \kiffj$.

In equations~(\ref{eq:kesji})-(\ref{eq:kiffj}), we have included
damping through $\nudamp=\nudampre+\nudampce$
and $\nudampi=\nudampri+\nudampci$, where
$\nudampre=(2e^2/3\me c^3)\omega^2$ and
$\nudampri=(Z^2\me/A\mpr)\nudampre$ are radiative damping rates
and $\nudampce=(\alpha_0\gff_\alpha/\nel\sigth)\nudampre$
and $\nudampci=(\me/A\mpr)\nudampce$ are collisional damping rates
(see Pavlov \etal~1995 and references therein).
Note that the natural width of the ion cyclotron resonance is
determined by the dimensionless damping rate
$\gamdampi\equiv\nudampi/\omega=\gamdampri+\gamdampci$, with
radiative and collisional contributions given by
\ba
\gamdampri & = & 5.18\times 10^{-9}\frac{Z^2}{A}
 \lp\frac{\hbar\omega}{\rm keV}\rp \label{eq:dampri} \\
\gamdampci & = & 8.42\times 10^{-8}\frac{Z^2}{A}\gff_\alpha
 \lp 1-e^{-\hbar\omega/kT}\rp \lp\frac{\mbox{10$^6$ K}}{T}\rp^{1/2}
 \lp\frac{\nion}{\mbox{10$^{24}$ cm$^{-3}$}}\rp
 \lp\frac{\rm keV}{\hbar\omega}\rp^2. \label{eq:dampci}
\ea
Also, the thermal width of the ion cyclotron line width is of order
\be
\gamdampth = \lp\frac{2kT}{A\mpr c^2}\rp^{1/2}
 = 1.46\times 10^{-3}A^{-1/2}\lp\frac{T}{\rm keV}\rp^{1/2}; \label{eq:dampth}
\ee
the Doppler broadening would presumably change the line profile
from the Lorentzian to the Voigt profile (e.g., Mihalas~1978).
Since $\gamdampri$, $\gamdampci$, $\gamdampth \ll 1$,
damping is negligible except very near resonance (see
Section~\ref{sec:opacity}).

\subsection{Behavior of Opacities and Ion Cyclotron Resonance}
\label{sec:opacity}

\begin{figure*}
\epsfxsize=15cm
\epsfysize=15cm
\epsffile{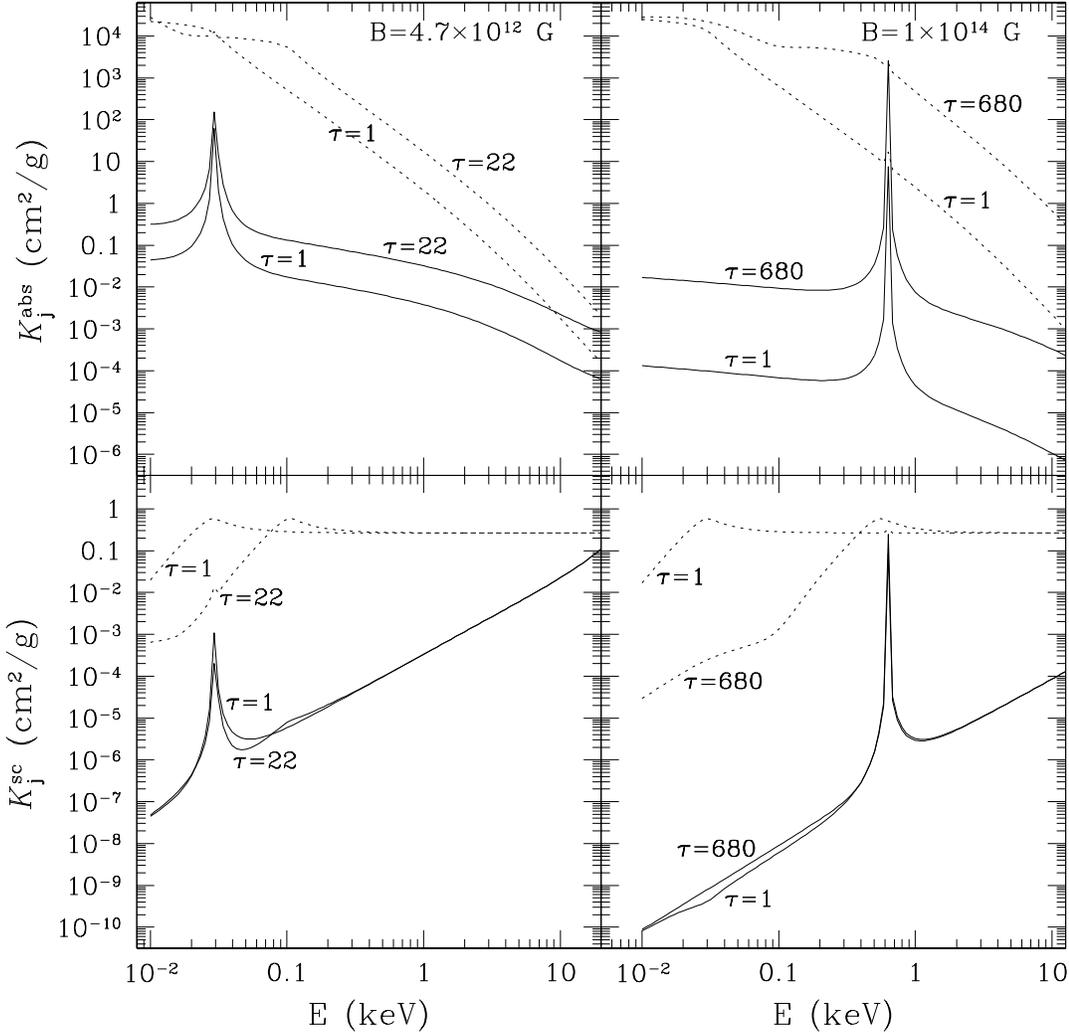}
\caption{Angle-averaged absorption opacities $\Kabsj$ (upper panels)
and scattering opacities $\Kscj$ (lower panels) as functions of
energy for magnetic fields $B=4.7\times 10^{12}$~G (left panels) and
$10^{14}$~G (right panels).  The densities and temperatures
are chosen to be representative of the hydrogen atmospheres
($Z=A=1$) studied in Section~\ref{sec:results} (see Fig.~\ref{fig:taunu}):
for $B=4.7\times 10^{12}$~G, Thomson depth $\tau=1$ corresponds
to $\rho=1$~g/cm$^3$ and $T=3.7\times 10^6$~K and $\tau=22$
corresponds to $\rho=13$~g/cm$^3$ and $T=5.9\times 10^6$~K;
for $B=10^{14}$~G, Thomson depth $\tau=1$ corresponds
to $\rho=1.1$~g/cm$^3$ and $T=3.4\times 10^6$~K and $\tau=680$
corresponds to $\rho=370$~g/cm$^3$ and $T=7.0\times 10^6$~K.
The solid lines are for the extraordinary mode ($j=1$) and
the dotted lines are for the ordinary mode ($j=2$).
We note that the nonmagnetic opacities are similar to the
ordinary mode opacities in terms of energy dependence,
and the variations in the ordinary mode opacities at low energies
($E\lo\hbar\omegap$) are due to $K_{z,j}$ [$\propto\omegap^2$;
see eq.~(\ref{eq:polarkz})].
}
\label{fig:ken}
\end{figure*}

Figure~\ref{fig:ken} shows the angle-averaged absorption
opacity $\Kabsj$ and scattering opacity $\Kscj=\sum_{i=1}^2\Kscji$
[see eqs.~(\ref{eq:kabsjavg})-(\ref{eq:kscjiavg})]
as functions of energy for magnetic fields
$B=4.7\times 10^{12}$~G and $10^{14}$~G at densities and
temperatures characteristic of NS atmospheres.
For angles $\thetab$ not too close to $0^\circ$ or $180^\circ$
(e.g., $10^\circ\la\thetab\la 170^\circ$), the unaveraged
opacities $\kscj$ and $\kabsj$ resemble $\Kscj$ and $\Kabsj$
in qualitative behavior, while exhibiting strong angle dependence
for $\thetab$ near $0^\circ$ or $180^\circ$.  The most prominent
feature in Fig.~\ref{fig:ken} is the ion cyclotron resonance at
$\hbar\omegabi=0.63\,B_{14}$~keV (for $Z=A=1$), where
$B_{14}=B/$(10$^{14}$~G), for the extraordinary mode.
Note that the cyclotron feature
does not possess the usual Voigt profile that characterizes
spectral lines; the feature is rather broad with
$\Delta E/E\sim 1+10^{-3}\,B_{14}^{-2}\rho_1$
[see eqs.~(\ref{eq:ionwidth})-(\ref{eq:ionwidth2}) below],
which is far wider than the natural or thermal width [see
eqs.~(\ref{eq:dampri})-(\ref{eq:dampth})].  Away from the resonance,
the extraordinary-mode opacities are suppressed relative to
the zero-field values by a factor $(\omega/\omegab)^2$
(see below).
For the ordinary mode, the opacities are similar to the
nonmagnetic values (see below) except at high densities/low frequencies,
i.e., $\vel\ga 1$, where the effects of $K_{z,j}$ become
important [see eq.~(\ref{eq:polarkz})].

The behavior of the opacities, especially the ion cyclotron feature,
shown in Fig.~\ref{fig:ken} can be understood as follows.
For the magnetic field strength and photon energy of interest
in this paper, $\hbar\omega\ll\hbar\omegab=11.6\,B_{12}$~keV,
the scattering and free-free absorption opacities can be
written as
\ba
\kscj=\frac{\nel\sigth}{\rho}\opbrace^j & , &
\kabsj=\frac{\alpha_0}{\rho}\opbrace^j,
\ea
with
\be
\opbrace^j\approx\lp\frac{1}{\uel}|e_+^j|^2+\frac{1}{\uel}|e_-^j|^2+|e_z^j|^2\rp
 + \frac{1}{M'^2}\lb\frac{1}{\lp 1-\sqrt{\uion}\rp^2+\gamdampi^2}|e_+^j|^2
 + \frac{1}{\lp 1+\sqrt{\uion}\rp^2}|e_-^j|^2 + |e_z^j|^2 \rb,
\label{eq:opbrace}
\ee
where $M'=(A\mpr/Z\me)Z^{-1/2}$ for scattering and
$M'=(A\mpr/Z\me)Z^{1/2}$ for absorption, and we have set
the free-free Gaunt factors to unity in $\kabsj$
and $A_\alpha=1$ in $\kscj$ for simplicity.
In the following, we shall consider photon energies such that
$|\uion-1|\ga\gamdampi$, $\gamdampth$ [see
eqs.~(\ref{eq:dampri})-(\ref{eq:dampth})], and we shall drop
the $\gamdampi^2$ term from equation~(\ref{eq:opbrace}).
Also for simplicity, we shall set $Z=A=1$.

For $\thetab$ not too small, i.e., $\sin^2\thetab/\cos\thetab\sim 1$,
the polarization parameter $\polarb$ [see eq.~(\ref{eq:polarb})]
satisfies $|\polarb|\gg 1$ except when $\omega$ is very close to
the ion cyclotron resonance (i.e., $|1-\uion-(1+\vel)/M|\la\uel^{-1/2}$)
or when $\hbar\omega\la\hbar\omegap^2/\omegab=0.07\rho_1B_{12}^{-1}$~eV.
In the limit of $|\polarb|\gg 1$, we find $K_1=-1/(2\polarb)$ and
$K_2=2\polarb$, and thus
\ba
|e_\pm^1|^2 = \frac{1}{2}\lp 1\mp\frac{1}{\polarb}\cos\thetab\rp & , &
|e_z^1|^2 = \frac{1}{\lp 2\polarb\rp^2}\sin^2\thetab
 \label{eq:polarvec1approx} \\
|e_\pm^2|^2 = \frac{1}{2}\cos^2\thetab\lp 1\pm\frac{1}{\polarb\cos\thetab}\rp
 & , & |e_z^2|^2 = \lb 1-\frac{1}{\lp 2\polarb\rp^2}\rb\sin^2\thetab,
 \label{eq:polarvec2approx}
\ea
where, for simplicity, we have taken the transverse approximation so that
$K_{z,j}\approx 0$.  Substituting these into
equation~(\ref{eq:opbrace}), we obtain, for $|\polarb|\gg 1$,
\ba
\opbrace^1 & \approx & \frac{1}{\uel}\lb 1+\frac{(1-\vel)^2
 \cos^2\thetab/\sin^2\thetab+\uion(1+\uion)}{(1-\uion)^2}\rb
 \label{eq:opbrace1} \\
\opbrace^2 & \approx & \sin^2\thetab+\frac{1}{\uel}\lb
 \lp\cos^2\thetab+\uion\sin^2\thetab\rp+\cos^2\thetab
 \frac{\uion(1+\uion)-(1-\vel)^2/\sin^2\thetab}{(1-\uion)^2}\rb
 \label{eq:opbrace2}.
\ea
Clearly, away from ion cyclotron resonance,
$\opbrace^1\propto\uel^{-1}=(\omega/\omegab)^2$, and thus the opacities
for the extraordinary mode are suppressed compared to the zero-field
values:
\ba
\ksc_1\propto\lp\frac{\omega}{\omegab}\rp^2 & , &
 \kabs_1\propto\frac{1}{\omegab^2\omega}\lp 1-e^{-\hbar\omega/kT}\rp.
\ea
At high densities/low frequencies, such that $\vel\ga 1$, these
are multiplied by a term of order $1+{\it O}(\vel^2)$; for
$\vel\gg 1$, we have $\ksc_1\propto\omegap^4/(\omegab^2\omega^2)$.
For the ordinary mode, $\opbrace^2\sim 1$, and thus the opacities
are approximately unchanged from the zero-field values.

Equations~(\ref{eq:opbrace1}) and (\ref{eq:opbrace2}) also apply
to photon energies near the ion cyclotron resonance as long as
$|1-\uion|\gg(1+\vel)/M$, so that $|\polarb|\gg 1$ (note that
we are considering typical angles where $|\cos\thetab|\sim 1$).
It is evident from equation~(\ref{eq:opbrace1}) that, for the
extraordinary mode, the opacities exhibit a broad peak/feature
around $\omegabi$, i.e., the term contained inside the brackets in
equation~(\ref{eq:opbrace1}) becomes significantly greater than
unity.  The width of the feature, $|1-\uion|$, is of order
\be
\Gamma_{\rm i}\equiv\left|\frac{\omega-\omegabi}{\omegabi}\right|
 \approx\lb 2+\lp1-\vel^{\rm res}\rp^2
 \frac{\cos^2\thetab}{\sin^2\thetab}\rb^{1/2}, \label{eq:ionwidth}
\ee
where
\be
\vel^{\rm res}=\frac{\omegap^2}{\omegabi^2} \approx
 \lp\frac{\rho}{\mbox{1 g/cm$^3$}}\rp
 \lp\frac{B}{4.6\times 10^{12}\mbox{ G}}\rp^{-2}. \label{eq:ionwidth2}
\ee
It is important to note that, although the natural and thermal
width of the ion cyclotron line are rather narrow
[see eqs.~(\ref{eq:dampri})-(\ref{eq:dampth})], the
cyclotron feature as described by equations~(\ref{eq:opbrace1})
and (\ref{eq:ionwidth}) is broad, i.e., the opacities are
significantly affected by the ion cyclotron resonance for
$0.3\la\omega/\omegabi\la 3$.  Also,
equation~(\ref{eq:opbrace2}) indicates that, although the
ion cyclotron resonance formally occurs for the ordinary
mode, its strength is diminished by the
$(\omega/\omegab)^2=(\me/\mpr)^2$ suppression.

Finally, we note that the presence of ions induces a mode-collapse
point very near $\omegabi$.  This can be seen
from equation~(\ref{eq:polarb}), which shows that $\polarb=0$ when
$1-\uion-(1+\vel)/M=0$ or $\omega=\omegabi[1+(1+\vel)/2M]$.
At this point, the two photon modes become degenerate (both are
circularly polarized), and the transport equation~(\ref{eq:rte0})
formally breaks down.  This ion-induced mode collapse is
independent of the well-known mode collapse due to vacuum
polarization (Pavlov \& Shibanov~1979; Ventura \etal~1979;
see Section~\ref{sec:discussion}).  Note, however, that the
ion-induced mode-collapse feature is very narrow
[$|\polarb|\ll 1$ requires $|1-\uion-(1+\vel)/M|\ll\uel^{-1/2}$],
so it is completely ``buried'' by the much wider and more
prominent ion cyclotron feature.

\setcounter{equation}{0}
\section{Numerical Method} \label{sec:numcomp}

\subsection{Magnetic Atmospheres with Full Radiative Transport}
\label{sec:numcompfull}

When presenting results of atmosphere models based on the
full transport equations, we consider only the case of
$\vecB$ perpendicular to the surface ($\Thetab=0$).
Thus the radiation intensity $\Inu^j$ depends on depth $\tau$,
energy $E=\hbar\omega=h\nu$, and angle $\theta$.

We construct a grid in Thomson depth, energy, and angle with intervals
that are equally spaced logarithmically from $10^{-4}$ to $\sim 10^3$
in depth and $10^{-2}$ to $\sim 10$~keV in energy and spaced every five
degrees in angle.  For typical calculations, six grid points are
used per decade in depth, twelve grid points are used per
decade in energy, and nineteen total grid points in angle.
We also increase the energy resolution near the ion cyclotron resonance.

The temperature profile is initially assumed to be the grey profile,
\be
T(\taur) = \Teff\lb\frac{3}{4}\lp\taur+\frac{2}{3}\rp\rb^{1/4}.
\label{eq:temptaur}
\ee
The Rosseland mean depth $\taur$ is given by
$d\taur = (\krtot/\keso)d\tau$, where the Rosseland mean opacity is
\be
\frac{1}{\krtot}
= \frac{3\pi}{4\sigsb T^3}\int_0^\infty \frac{1}{2}\sum_j\mfpj
\frac{\partial\Bnu}{\partial T}d\nu \label{eq:kr}
\ee
and $\mfpj$ is given by equation~(\ref{eq:defnmfpj}).
Since $\mfpj$ and $\krtot$ are functions of $\rho$ and $T$
themselves, the profile~(\ref{eq:temptaur}) must be constructed
by an iterative process: the temperature profile is first taken
to be $T(\tau) = \Teff[3/4(\tau+2/3)]^{1/4}$; the density profile,
and hence $\ktot$ and $\krtot$, are then determined,
which allows the calculation of $\taur(\tau)$; using
equation~(\ref{eq:temptaur}) with this $\taur$, a revised
temperature profile is obtained; the process is repeated
until the new $\taur(\tau)$ is not significantly different
from the previous $\taur(\tau)$.
To achieve $|\Delta\tau/\tau|$ and $|\Delta\taur/\taur| < 0.1 \%$,
three iterations are required.

To construct self-consistent atmosphere models requires
successive global iterations, where the temperature profile
$T(\tau)$ is adjusted from the previous iteration
to satisfy radiative equilibrium.  At each global iteration,
the profiles $T(\tau)$ and $\rho(\tau)$ are considered fixed.
We solve the RTE~(\ref{eq:rte1}) [see also
eqs.~(\ref{eq:rtestandard})-(\ref{eq:sourceapprox})],
together with the boundary conditions
[see eqs.~(\ref{eq:bc1})-(\ref{eq:bc2})] by the finite difference
scheme described in Mihalas~(1978).  This results in a set of equations
which have the form of a tridiagonal matrix that
can be solved by the Feautrier procedure of forward-elimination
and back-substitution.
We have also implemented an improved Feautrier method to
reduce errors arising from machine precision as suggested by
Rybicki \& Hummer~(1991).
Note that, in the presence of scattering, the source function
$\Snu^j$ [which involves mode coupling; see eq.~(\ref{eq:source})]
is not known prior to solving the RTE.  We adopt the strategy
that, for each global iteration, $\Snu^j$ is calculated using
the solution for $\meani$ from the previous global iteration,
with $\Snu^j=\Bnu/2$ for the first global iteration.
The full RTE is solved with this source function to obtain $\meani$,
and the specific flux is calculated from finite differencing
equation~(\ref{eq:rte1b}).
Note that, unless the temperature correction from the previous
iteration is relatively small, so that the source function
calculated from the previous $\meani$ approximates the true
current source function, the global iteration may not converge.\footnote{
An alternative is to calculate $\Snu^j$ iteratively within
each global iteration.  The source function during each global
iteration is first taken to be $\Snu^j=\Bnu/2$, and
thus the two modes are initially decoupled.  The RTE is then
solved for each mode $j$.  The resulting $\meani$
is used to determine a revised source function,
which can then be reinserted into the RTE.  This procedure is
repeated until the the source function converges,
e.g., $|\Delta \Snu^j/\Snu^j| < 5\%$.  Obviously, this method
requires longer computation time.  In addition, we have
implemented a procedure that reformulates the RTE using
Eddington factors (see, e.g., Rybicki \& Lightman~1979
for the analogous nonmagnetic RTE using Eddington factors), but this does
not lead to significant reduction in computation time.
\label{foot:sourceitr}}

The current solution $\meani$ does not, in general,
satisfy the radiative equilibrium constraint nor does it yield a
constant flux at every depth [see eq.~(\ref{eq:radeq}) and
(\ref{eq:fluxconst}), respectively].  To satisfy both
conditions requires correcting the temperature profile.
We use a variation of the Uns\"{o}ld-Lucy
temperature correction method as described in Mihalas~(1978)
but modified to account for full radiative transfer by two propagation
modes in a magnetic medium.  The radiative equilibrium
constraint given by equation~(\ref{eq:radeq}) can be written as
\be
\frac{dF_z}{d\tau}=c\lp\frac{\kj}{\keso}u-\frac{\kp}{\keso}\upl\rp,
 \label{eq:radeqz}
\ee
where $u=\sum_{j=1}^2\int d\nu\,\unuj$ is the total radiation
energy density, $\upl=aT^4=(4\sigsb/c)T^4$ is the
blackbody energy density, and $\kj$ and $\kp$ are the
absorption mean opacity and Planck mean opacity defined by
\ba
\kj & \equiv & \frac{2}{cu}\sum_{j=1}^2 \int d\nu\int_+ d\veck\,
 \kabsj(\veck)\meani(\veck) \label{eq:kj} \\
\kp & \equiv & \frac{1}{4\pi\upl} \sum_{j=1}^2 \int d\nu\int_+ d\veck\,
 \kabsj(\veck)\unup, \label{eq:kp}
\ea
respectively.  On the other hand, equation~(\ref{eq:rte1b})
can be integrated out formally to give $\meani$ in terms
of $\meanf$, and we then have
\be
cu(\tau) = 2\sum_{j=1}^2\int d\nu\int_+ d\veck\,\meani(\tau)
 = \int_0^\tau d\tau'\frac{\kf(\tau')}{\keso}F_z(\tau')+2F_z(0),
 \label{eq:endenstot}
\ee
where we have used $cu(0)\approx 2F_z(0)$, $F_z(\tau)$ is the
total flux at depth $\tau$ [see eq.~(\ref{eq:fluxconst})], and
$\kf(\tau)$ is the flux mean opacity defined by
\be
\kf \equiv \frac{2}{F_z} \sum_{j=1}^2 \int d\nu\int_+ d\veck\,
 \frac{1}{\mu}\ktotj(\veck)\meanf(\veck). \label{eq:kf}
\ee
Combining equations~(\ref{eq:radeqz}) and (\ref{eq:endenstot})
gives an expression for $\upl[T(\tau)]$ in terms of $F_z(\tau)$.
The desired temperature correction $\Delta T(\tau)$ needed to
satisfy radiative equilibrium is then
\be
\Delta T(\tau) \approx \frac{1}{16\sigsb T(\tau)^3}
\left\{-\frac{\keso}{\kp(\tau)}\frac{d\lb\Delta F_z(\tau)\rb}{d\tau} +
\frac{\kj(\tau)}{\kp(\tau)}\lb \int_0^\tau d\tau'\frac{\kf(\tau')}{\keso}
 \Delta F_z(\tau') + 2\Delta F_z(0)\rb\right\},
\label{eq:tempcor0}
\ee
where $\Delta F_z(\tau) = \sigsb\Teff^4 - F_z(\tau)$.
Using equation~(\ref{eq:radeqz}) to replace $dF_z/d\tau$
in equation~(\ref{eq:tempcor0}), we obtain
\be
\Delta T(\tau) \approx \frac{1}{16\sigsb T(\tau)^3}
\left\{\frac{c}{\kp(\tau)}\lb\kj(\tau)u(\tau)-\kp(\tau)\upl(\tau)\rb+
\frac{\kj(\tau)}{\kp(\tau)}\lb \int_0^\tau d\tau'\frac{\kf(\tau')}{\keso}
 \Delta F_z(\tau') + 2\Delta F_z(0)\rb\right\}.
\label{eq:tempcor}
\ee
Note that the first term in equation~(\ref{eq:tempcor}),
which is $\propto(\kj u-\kp\upl)$, corresponds to the
temperature correction in the $\Lambda$-iteration procedure
$\Delta T_\Lambda=(\kj u-\kp\upl)(\partial\upl/\partial T)^{-1}$
and is important near the surface but is small in the deeper
layers where the atmosphere approaches a blackbody, while the
remaining terms provide corrections in these deeper layers.
In practice, we find that the Uns\"{o}ld-Lucy temperature
correction method as defined by equation~(\ref{eq:tempcor})
tends to over-correct;  therefore we only use $\sim$70-90\%
of the temperature correction.
The process of determining the radiation intensity from
the RTE for a given temperature profile, estimating and
applying the temperature correction, and then recalculating
the radiation intensity is repeated until convergence
of the solution is achieved.
We note here that, to decrease the number of iterations
required for convergence and the total computation time,
we can use as the initial temperature profile the result
obtained from the diffusion calculation (see Section~\ref{sec:numcompd});
such an initial temperature profile would already be close
to radiative equilibrium.

\begin{figure*}
\epsfxsize=15cm
\epsfysize=15cm
\epsffile{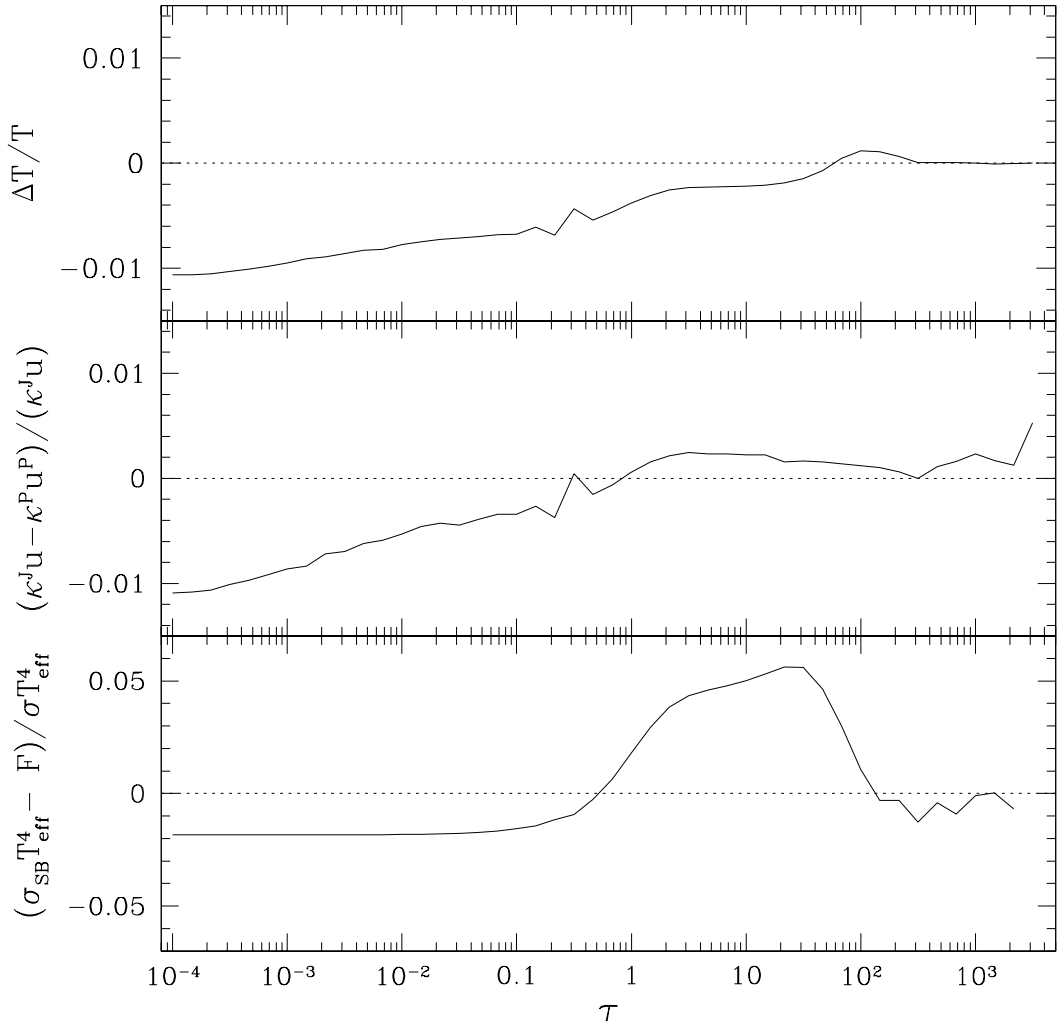}
\caption{Convergence tests after $\sim$20 global iterations (temperature
corrections) for a fully ionized hydrogen atmosphere model with
$\Teff=5\times 10^6$~K, $B=10^{14}$~G, and $\Thetab=0$.
Fractional temperature correction (upper panel), deviation from
radiative equilibrium [middle panel; see
eqs.~(\ref{eq:radeqz})-(\ref{eq:kp})], and deviation from constant
total flux (lower panel) are plotted as functions of Thomson depth $\tau$.
The larger deviations around $\tau\sim 1-100$ in the lower
panel are due to the extraordinary-mode photons decoupling
from the matter at $\tau\sim 100$ (see Fig.~\ref{fig:taunu});
this region has the slowest convergence, but the deviations
can be reduced further with additional iterations.}
\label{fig:constraints}
\end{figure*}

Three criteria are used to indicate convergence.  The first
is that the temperature correction between successive global
iterations becomes small, e.g., $|\Delta T/T|\la 1\%$.
The second is to check that radiative equilibrium
[see eq.~(\ref{eq:radeq})] is sufficiently satisfied, or
$(\kj u-\kp\upl)/(\kj u)$ is sufficiently small
[see eq.~(\ref{eq:radeqz})].  The third criterion is that
$F_z$ is sufficiently close to $\sigsb\Teff^4$
[see eq.~(\ref{eq:fluxconst})].  Figure~\ref{fig:constraints}
shows an example of these convergence tests after
about twenty global iterations for our full radiative
transport model of a fully ionized, pure hydrogen atmosphere
with $\Teff=5\times 10^6$~K, $B=10^{14}$~G, and $\Thetab=0$.
Further iterations reduce the deviations, though the
convergence is slower since the temperature corrections are
already small.  The numerical results presented in Section~\ref{sec:results}
have all reached the convergence level similar to Fig.~\ref{fig:constraints}.

\subsection{Magnetic Atmospheres with Diffusion Approximation}
\label{sec:numcompd}

Using the diffusion approximation of radiative transport
(see Section~\ref{sec:rted}), we construct magnetic atmosphere
models for general $\Thetab$.  We follow a similar procedure
as the full radiative transfer method described in
Section~\ref{sec:numcompfull}.  We solve the approximate
RTE~(\ref{eq:rte1d}) with boundary conditions given by
equations~(\ref{eq:bc1d}) and (\ref{eq:bc2d}) using the
Feautrier method, which involves only depth and photon
energy grid points.  The modified Uns\"{o}ld-Lucy temperature
correction can be derived using
the specific flux perpendicular to the stellar surface,
$F_{\nu,z}^j\approx-c\mfpj(\partial\unuj/\partial z)$, and
equation~(\ref{eq:radeqd}).  The resulting $\Delta T(\tau)$ takes the
same form as in equations~(\ref{eq:tempcor0}) and (\ref{eq:tempcor}),
except that the mean opacities $\kj$, $\kp$, and $\kf$ are
given by
\ba
\kj & \equiv & \frac{1}{u}\sum_{j=1}^2\int d\nu\,\Kabsj\unuj \\
\kp & \equiv & \frac{1}{2\upl}\sum_{j=1}^2\int d\nu\,\Kabsj\unup \\
\kf & \equiv & \frac{1}{F_z}\sum_{j=1}^2\int d\nu\,\frac{1}{\rho\mfpj}\Fnu^j,
\ea
respectively.  The expressions for $\kj$ and $\kp$
are essentially the same as in Section~\ref{sec:numcompfull}.
Atmosphere models for $B\la 10^{13}$~G
based on the diffusion approximation have been constructed by
Shibanov \etal~(1992); Pavlov \etal~(1995) using a different iteration
scheme, i.e., $\Lambda$-iteration.

\subsection{Nonmagnetic Atmospheres}
\label{sec:numcompnonmag}

For comparison, we also construct nonmagnetic atmosphere models
by solving the standard RTE (both full transport and in the
diffusion approximation) using a similar method as described in
Section~\ref{sec:numcompfull} and Section~\ref{sec:numcompd}.
In this case, the opacities are just due to
nonmagnetic electron scattering and free-free absorption.
We implement the RTE with Eddington factors to
more accurately determine the source function.  We use the
standard Feautrier method
(with improvements described in Rybicki \& Hummer~1991)
and the standard Uns\"{o}ld-Lucy temperature correction scheme
described in Mihalas~(1978).

\setcounter{equation}{0}
\section{Numerical Results} \label{sec:results}

\subsection{Atmosphere Structure} \label{sec:atmstructure}

\begin{figure*}
\epsfxsize=15cm
\epsfysize=15cm
\epsffile{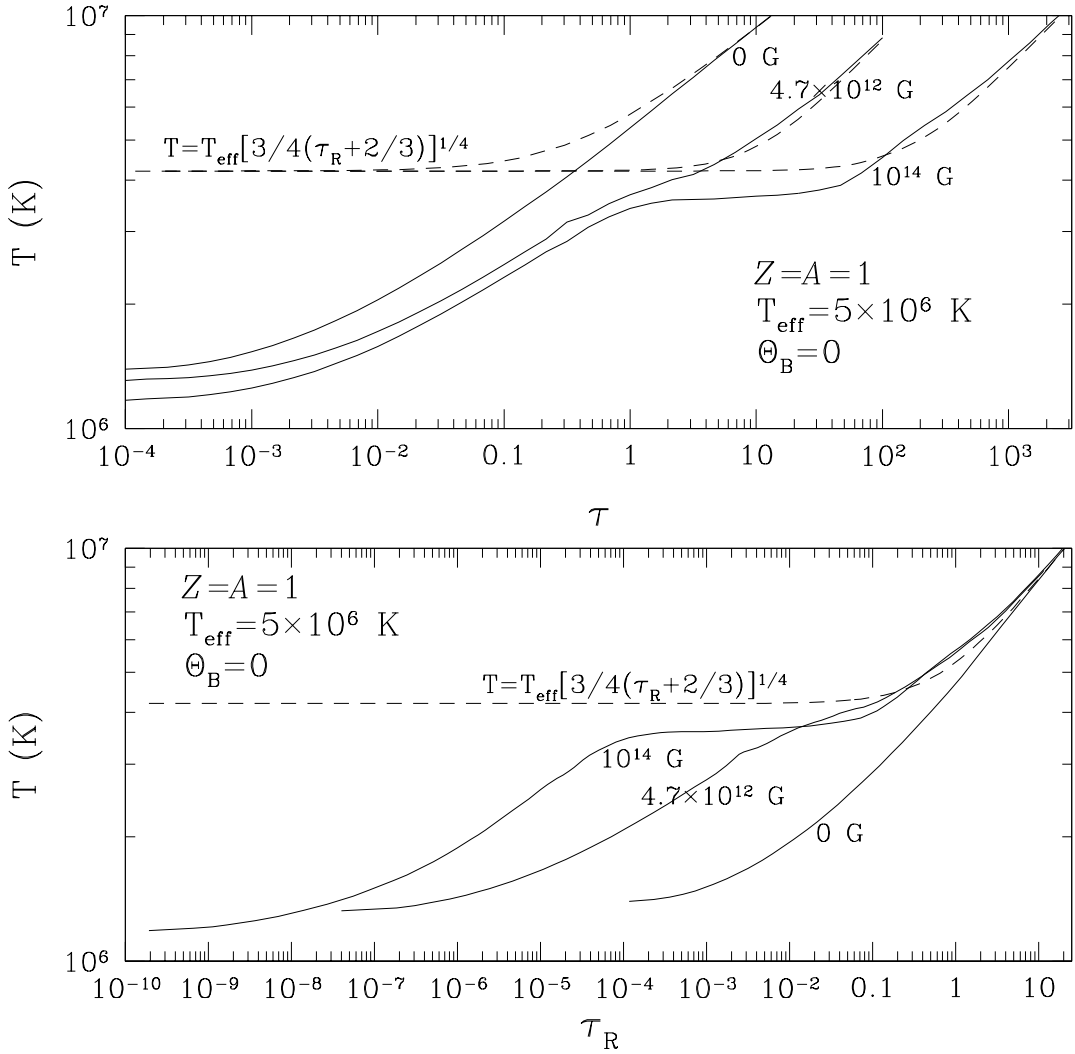}
\caption{Temperature $T$ as a function of Thomson depth $\tau$
(upper panel) and Rosseland mean depth $\taur$ (lower panel)
for fully ionized, pure hydrogen atmospheres with
$\Teff=5\times 10^6$~K, and magnetic fields
$B=0$, $4.7\times 10^{12}$~G, $10^{14}$~G and $\Thetab=0$.
The dashed lines show the grey temperature profile
given by equation~(\ref{eq:temptaur}).}
\label{fig:taur}
\end{figure*}

Figure~\ref{fig:taur} shows the temperature profile as a function
of Thomson depth $\tau$ and Rosseland depth $\tau_R$ for
hydrogen atmospheres with $\Teff=5\times 10^6$~K, magnetic
fields $B=4.7\times 10^{12}$~G and $10^{14}$~G,
and the magnetic field oriented perpendicular to the surface
($\Thetab=0$). For comparison, Fig.~\ref{fig:taur} also shows
the temperature profile of our nonmagnetic atmosphere model and
the grey profile given by equation~(\ref{eq:temptaur}).
Note that, since the Rosseland mean opacity [eq.~(\ref{eq:kr})]
is dominated by the (small) extraordinary-mode opacity, the
Rosseland depth $\tau_R(\tau)$
is significantly smaller at high fields than at $B=0$.
It is evident that all models approach the grey profile
at $\tau_R\gg 1$, although this occurs deeper in the
atmosphere for models with higher magnetic fields.
At small depths of $\tau\la 1$, the gas temperature is smaller, by a factor
of a few, than predicted by the grey profile. This is because,
for $\tau\la 1$, the extraordinary-mode photons have already decoupled from
the matter, and the temperature is primarily determined by the ordinary-mode
absorption opacity, which is strongly nongrey ($\kabs_2\propto
\nu^{-3}$ approximately). The plateau in the $T(\tau)$ profile at
$\tau\sim 1-100$ in the $B=10^{14}$~G model arises because the
extraordinary-mode photons decouple from the matter at $\tau\sim 100$
(see Section~\ref{sec:opticaldepth}); such a plateau is more distinctive
for high magnetic fields
and becomes less pronounced for smaller magnetic fields.

\begin{figure*}
\epsfxsize=15cm
\epsfysize=15cm
\epsffile{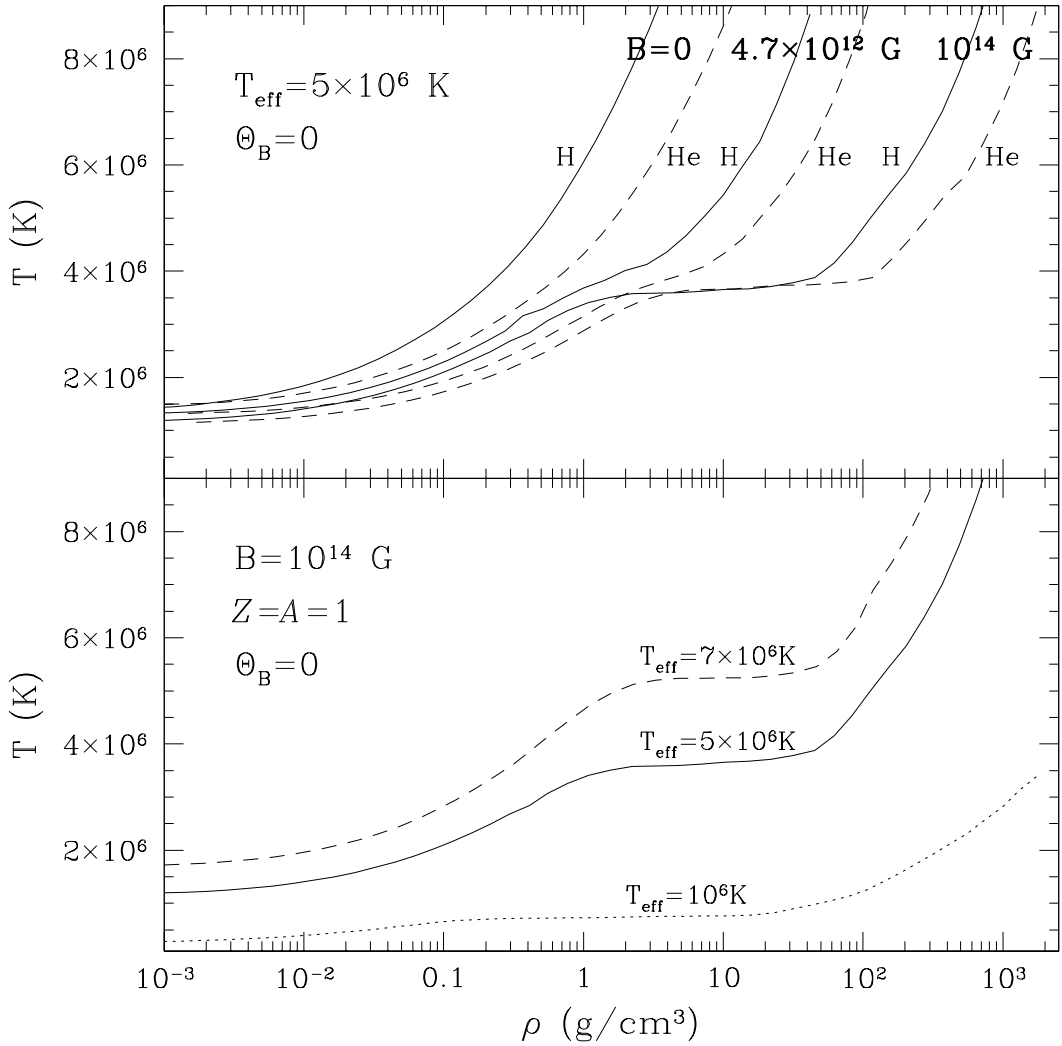}
\caption{Temperature as a function of density for the various
fully ionized atmosphere models considered in Section~\ref{sec:spectra}.
The upper panel shows the H and He models at $B=0$, $4.7\times 10^{12}$~G,
and $10^{14}$~G, all for $\Teff=5\times 10^6$~K and $\Thetab=0$.
The lower panel shows the H models at $B=10^{14}$~G for
$\Teff=10^6$~K, $5\times 10^6$~K, and $7\times 10^6$~K.}
\label{fig:tempdens}
\end{figure*}

Figure~\ref{fig:tempdens} shows the temperature versus density
profiles for various atmosphere models considered in this paper
(see Section~\ref{sec:spectra} for the spectral properties of
these models). The plateau in temperature for the $B=10^{14}$
models correspond to the similar feature shown in Fig.~\ref{fig:taur}.
Note that, in all models, the electrons are nondegenerate at the
relevant densities, i.e., $T\gg T_F$ [see eq.~(\ref{eq:tempfermi})].
Thus, even for high (quantizing) magnetic fields, the equation
of state deviates from the classical ideal gas law by less than
a few percent (see Section~\ref{sec:eos}).  At a given $B$, the
temperature profile, $T(\tau)$, of the He atmosphere
is very close to that of the H atmosphere (see Fig.~\ref{fig:taur});
this is expected since, apart from the ion cyclotron resonances, the H and
He opacities have the same frequency dependence, and differ only by a
factor of $\sim 2$ in magnitude. With $P\simeq (1+Z)\rho kT/A$, we see
that the He atmosphere has a higher density, by a factor of $\sim 16/3$,
than the H atmosphere at the same temperature (see Fig.~\ref{fig:tempdens}).

\subsection{Photon-Matter Decoupling Region}
\label{sec:opticaldepth}

\begin{figure*}
\epsfxsize=15cm
\epsfysize=15cm
\epsffile{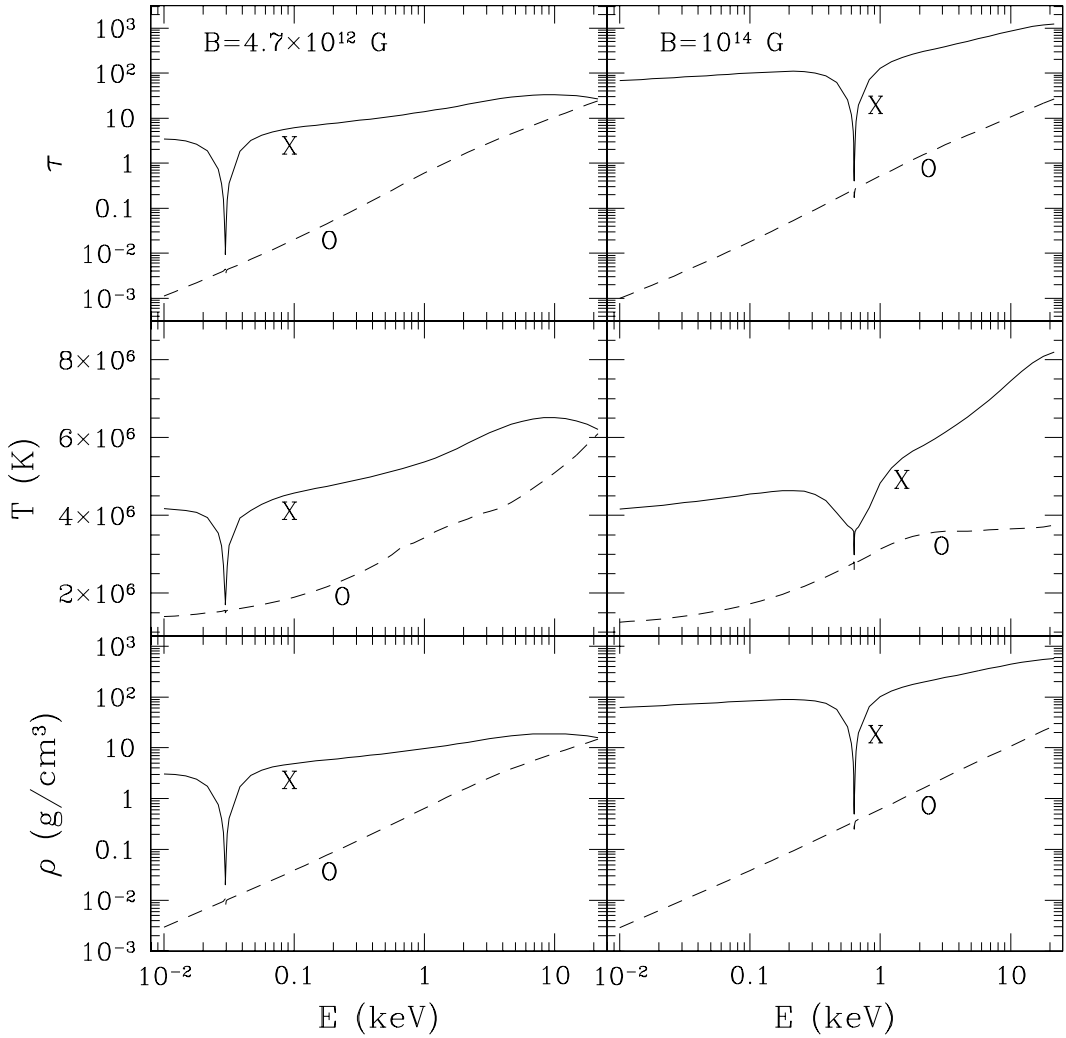}
\caption{The upper panels show the Thomson depth at which
photons decouple from the matter [effective optical depth
$\tau_{\nu j}^{\rm eff}=1$; see eq.~(\ref{eq:taunu})]
as a function of energy for the $\Teff=5\times 10^6$~K,
$\Thetab=0$, and $B=4.7\times 10^{12}$~G (left panels)
and $10^{14}$~G (right panels) atmosphere models.
Also plotted are the local temperature $T$ (middle panels)
and density $\rho$ (lower panels) at these Thomson depths.
The solid lines are for the extraordinary mode ($j=1$), and the
dashed lines are for the ordinary mode ($j=2$).}
\label{fig:taunu}
\end{figure*}

To better understand the atmosphere structure and the emergent
spectra (see Section~\ref{sec:spectra}), it is useful to develop
a rough idea of the physical conditions in the region of the
atmosphere where photons
of different modes and energies decouple from the matter.
Because of the angle dependence of the opacities, the exact location of
the ``decoupling layer'' is ambiguous. One possible approach
is to use the averaged opacities defined in Section~\ref{sec:rted};
the effective opacity for photon-matter energy exchange is
$K_j^{\rm eff}=(\Kabsj\bar\Ktotj)^{1/2}$ (see Rybicki \& Lightman~1979),
where $\bar\Ktotj\equiv (\rho\mfpj)^{-1}$ and
$\mfpj$ and $\Kabsj$ are given by
equations~(\ref{eq:defnmfpj}) and (\ref{eq:kabsjavg}), respectively.
In fact, if mode coupling due to scattering is neglected,
equation~(\ref{eq:rte1d}) can be approximately integrated out to yield
$\unuj(y=0)\propto \unup(y\sim 1/K_j^{\rm eff})$, i.e.,
the emergent radiation density (of mode $j$) is determined by the blackbody
radiation density evaluated at the depth where the column density $y\sim
1/K_j^{\rm eff}$. We therefore define the effective optical depth
\be
\tau_{\nu j}^{\rm eff}(\tau)=\frac{1}{\keso}\int_0^\tau d\tau'
\lp{\Kabsj\bar\Ktotj}\rp^{1/2}=\frac{1}{\keso}\int_0^\tau d\tau'
 \lp\frac{\Kabsj}{\rho\mfpj}\rp^{1/2}, \label{eq:taunu}
\ee
so that photons of mode $j$ and frequency $\nu$ decouple
from the matter at $\tau_{\nu j}^{\rm eff}\sim 1$.
Figure~\ref{fig:taunu} shows the Thomson depth and local
temperature and density as functions
of photon energy at the decoupling layer
for the H atmosphere model with
$\Teff=5\times 10^6$~K, $\Thetab=0$, and $B=4.7\times 10^{12}$~G
and $10^{14}$~G (see also Fig.~\ref{fig:taur}).
Note that the effective depth defined in equation~(\ref{eq:taunu})
errs for photons propagating along the magnetic field since the opacities
are generally lower near $\thetab=0$; hence, these photons decouple
from the matter at deeper layers than those indicated in Fig.~\ref{fig:taunu}.
Nevertheless, Fig.~\ref{fig:taunu} gives the typical
conditions where the observed photons of energy $E$ are generated.
It is clear that the extraordinary-mode photons can emerge
from deep in the atmosphere where plasma effects on the opacities
are not negligible (see Section~\ref{sec:opacity0} and Fig.~\ref{fig:ken}).
Photons with energies near the ion cyclotron resonance
$E_{Bi}=0.63\,B_{14}$~keV decouple in the lower temperature region;
this gives rise to the absorption feature in the emergent radiation
spectrum (see Section~\ref{sec:spectra}).

\subsection{Spectra} \label{sec:spectra}

\begin{figure*}
\epsfxsize=15cm
\epsfysize=15cm
\epsffile{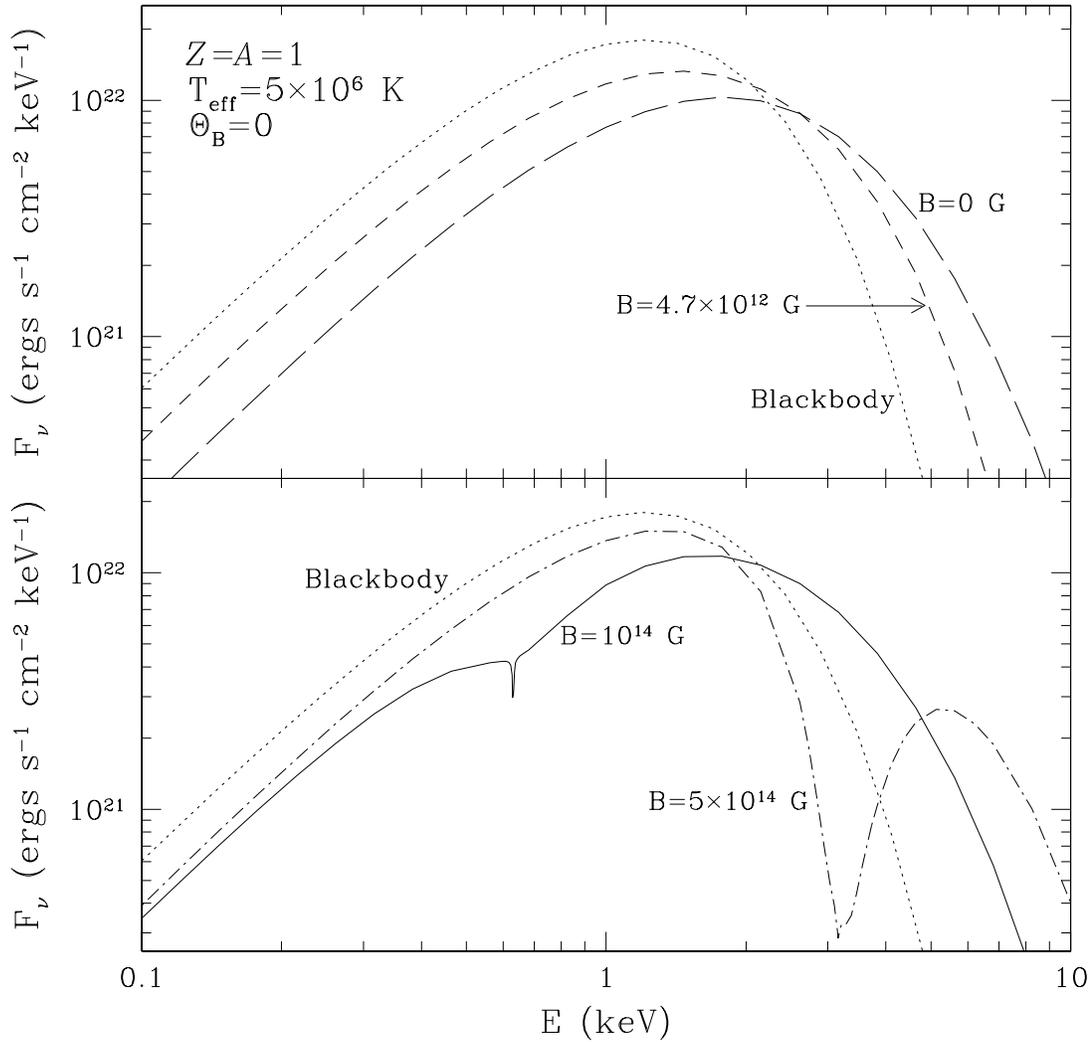}
\caption{Spectra of neutron star atmospheres with
fully ionized hydrogen at $\Teff = 5\times 10^6$~K and $\Thetab=0$.
The dot-dashed line is for the $B=5\times 10^{14}$~G atmosphere,
the solid line is for the $B=10^{14}$~G atmosphere,
the short-dashed line is for the $B=4.7\times 10^{12}$~G atmosphere,
the long-dashed line is for the nonmagnetic atmosphere,
and the dotted lines are for a blackbody with $T = 5\times 10^6$~K.
At $B=5\times 10^{14}$~G, the $\Ebi=3.15$~keV ion cyclotron feature
(from $\sim$2 to $\sim$5~keV) has an equivalent width $\sim$1.9~keV,
while at $B=10^{14}$~G, the $\Ebi=0.63$~keV feature (from $\sim$0.4
to $\sim$1~keV) has an equivalent width $\sim$90~eV.  Note that
the extremely narrow natural/thermal lines at $\Ebi$ have negligible
equivalent widths.}
\label{fig:spectrum5h}
\end{figure*}

Figure~\ref{fig:spectrum5h} shows the spectra of the hydrogen atmospheres
with $\Teff=5\times 10^6$~K, $\Thetab=0$, and magnetic fields of
$B=4.7\times 10^{12}$, $10^{14}$, and $5\times 10^{14}$~G.
Also plotted are the spectrum of our nonmagnetic fully ionized H
atmosphere model and the blackbody spectrum at $T=5\times 10^6$~K.
The spectra clearly show a significantly harder high energy tail and
a depletion of low energy photons relative to the blackbody.  For example,
in the case of $B=10^{14}$~G, the flux at $E=5$~keV is an order of magnitude
larger than the corresponding blackbody flux. At a given energy $E$
(and for the same $\Teff$), the flux depends on the field strength in a
non-monotonic fashion for the models depicted in Fig.~\ref{fig:spectrum5h}.
For example, at $E=7$~keV, the flux increases in the order of
$B=4.7\times 10^{12},~10^{14},~0,~5\times 10^{14}$~G.
The hard tails were already noted in previous studies of NS atmospheres with
$B\la 10^{13}$~G (e.g., Pavlov \etal~1995;
see also references cited in Section~\ref{sec:intro}),
and they arise because high energy photons have smaller opacities
and thus decouple from deeper, hotter layers (see Fig.~\ref{fig:taunu}).
It is evident that fitting the high energy tail of observations with a
blackbody will yield overestimates of the surface temperature of the NS.

\begin{figure*}
\epsfxsize=15cm
\epsfysize=15cm
\epsffile{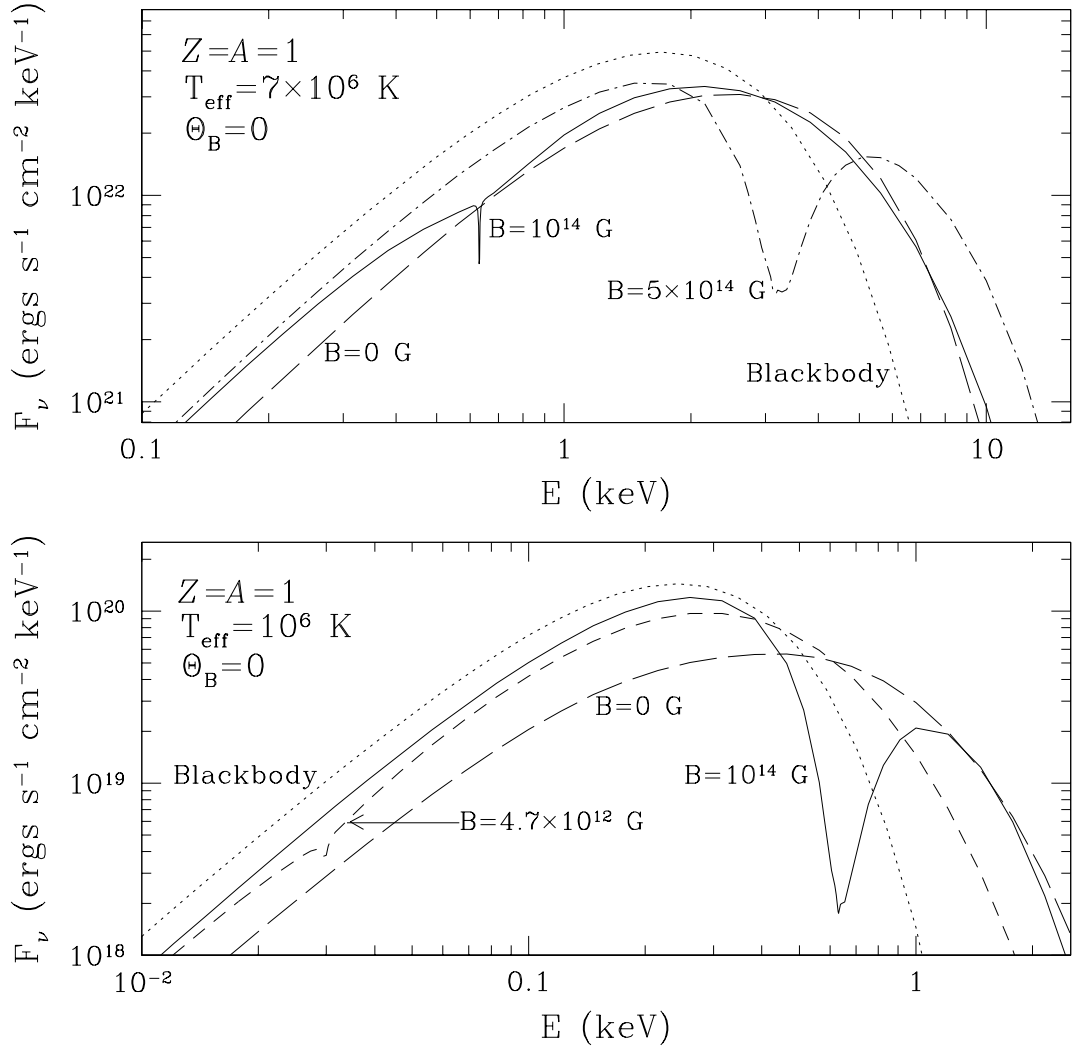}
\caption{Spectra of neutron star atmospheres with
fully ionized hydrogen at $\Teff=7\times 10^6$ (upper
panel) and $10^6$~K (lower panel) and $\Thetab=0$.
The dot-dashed line is for the $B=5\times 10^{14}$~G atmosphere,
the solid lines are for the $B=10^{14}$~G atmosphere,
the short-dashed line is for the $B=4.7\times 10^{12}$~G atmosphere,
the long-dashed lines are for the nonmagnetic atmosphere,
and the dotted lines are for a blackbody with $T = 7\times 10^6$~K
and $10^6$~K.
Note that the ion cyclotron resonance features are more pronounced
when $\Ebi\go 3\,k\Teff$:
at $B=5\times 10^{14}$~G and $\Teff=7\times 10^6$~K,
the $\Ebi=3.15$~keV feature
(from $\sim$2 to $\sim$5~keV) has an equivalent width $\sim$1.8~keV,
while at $B=10^{14}$~G and $\Teff=10^6$~K, the $\Ebi=0.63$~keV
feature (from $\sim$0.4 to $\sim$1~keV) has an equivalent width
$\sim$0.4~keV.  Also note that the extremely narrow natural/thermal
lines at $\Ebi$ have negligible equivalent widths.
}
\label{fig:spectrumtemp}
\end{figure*}

\begin{figure*}
\epsfxsize=15cm
\epsfysize=15cm
\epsffile{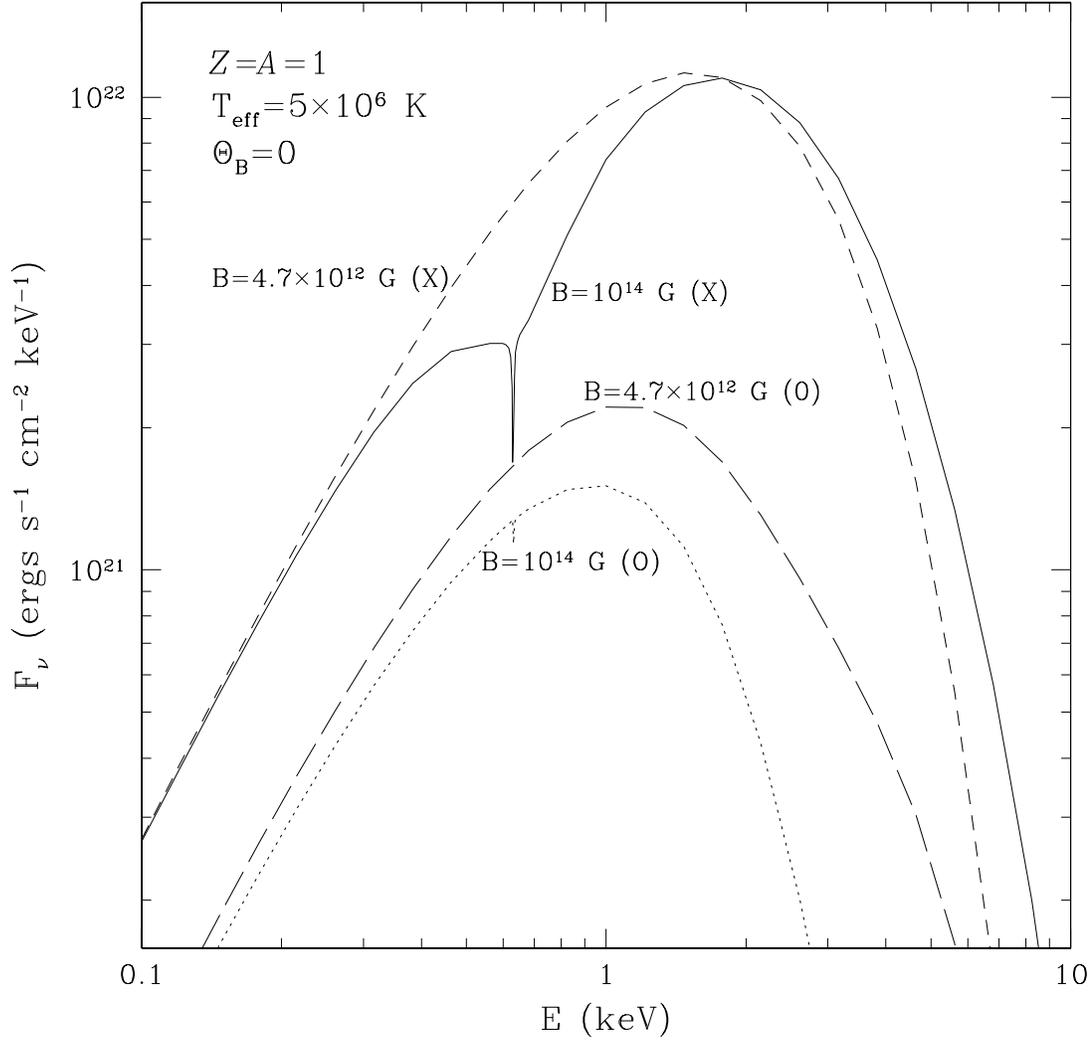}
\caption{Spectra of neutron star atmospheres with fully ionized
hydrogen at $\Teff=5\times 10^6$~K and $\Thetab=0$.
The solid and dotted lines are for the $B=10^{14}$~G extraordinary
mode ($j=1$) and ordinary mode ($j=2$), respectively, and
the short and long-dashed lines are for the $B=4.7\times 10^{12}$~G
extraordinary and ordinary modes, respectively.}
\label{fig:spectrumpolar}
\end{figure*}

\begin{figure*}
\epsfxsize=15cm
\epsfysize=15cm
\epsffile{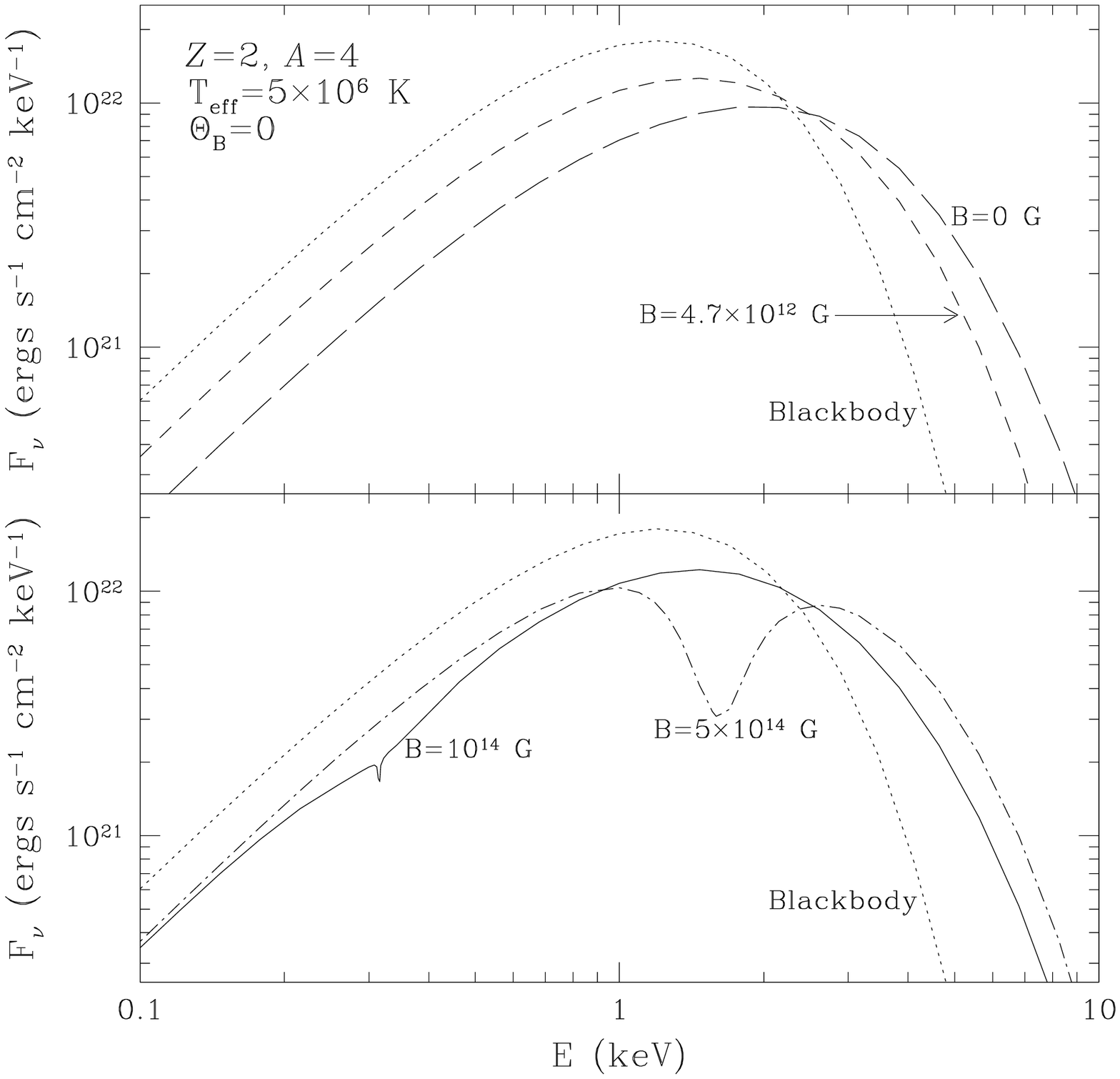}
\caption{Spectra of neutron star atmospheres with
fully ionized helium at $\Teff = 5\times 10^6$~K and $\Thetab=0$.
The dot-dashed line is for the $B=5\times 10^{14}$~G atmosphere,
the solid line is for the $B=10^{14}$~G atmosphere,
the short-dashed line is for the $B=4.7\times 10^{12}$~G atmosphere,
the long-dashed line is for the nonmagnetic atmosphere,
and the dotted lines are for a blackbody with $T = 5\times 10^6$~K.
At $B=5\times 10^{14}$~G, the $\Ebi=1.58$~keV ion cyclotron feature
(from $\sim$1 to $\sim$2~keV) has an equivalent width $\sim$0.51~keV,
while at $B=10^{14}$~G, the $\Ebi=0.32$~keV feature (from $\sim$0.2
to $\sim$0.4~keV) has an equivalent width $\sim$10~eV.}
\label{fig:spectrumcomp}
\end{figure*}

\begin{figure*}
\epsfxsize=15cm
\epsfysize=15cm
\epsffile{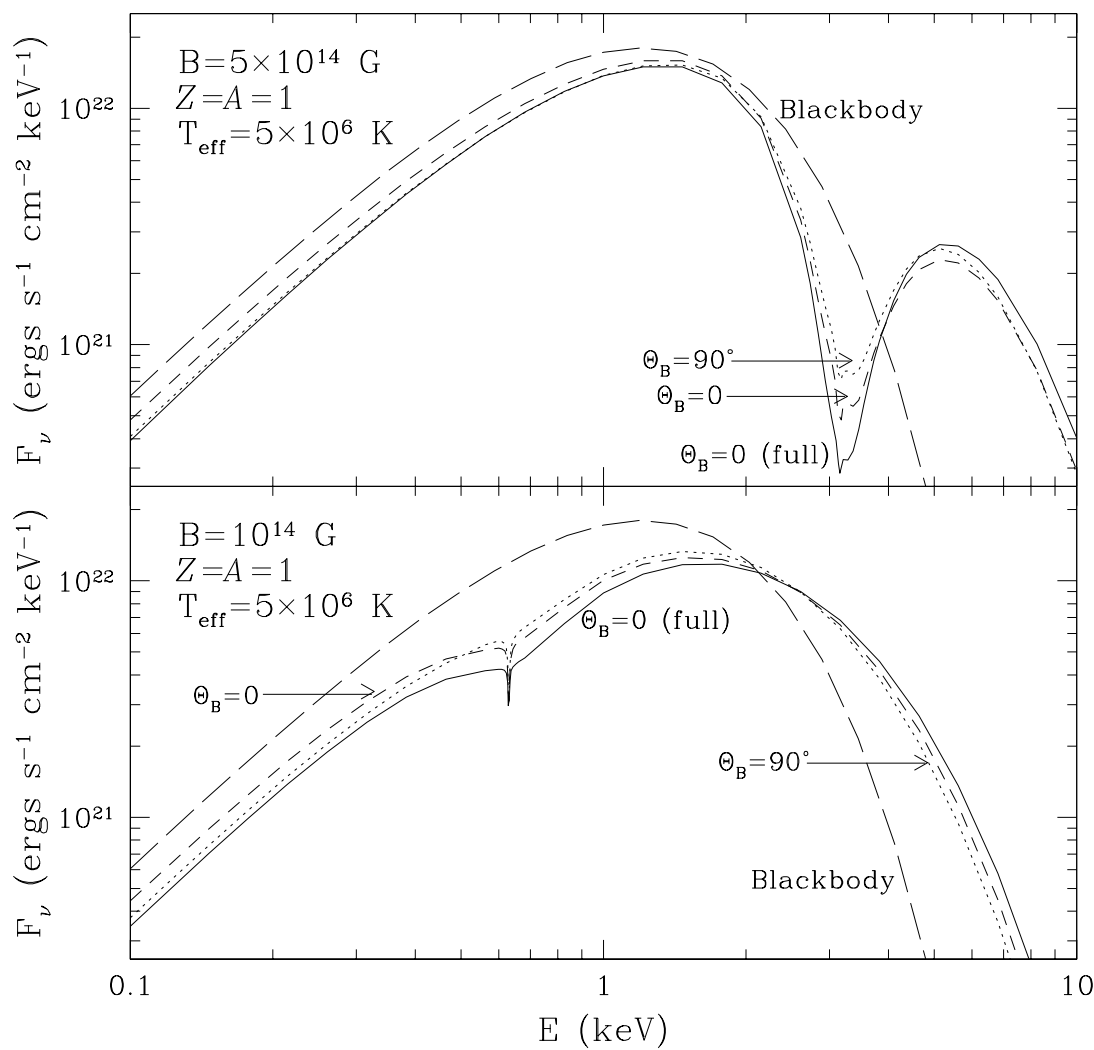}
\caption{Spectra of neutron star atmospheres with fully ionized
hydrogen at $B=5\times 10^{14}$~G (upper panel) and
$10^{14}$~G (lower panel) and $\Teff=5\times 10^6$~K.
The solid lines are for $\Thetab=0$ (full radiative transport),
the short-dashed lines are for $\Thetab=0$ (diffusion),
the dotted lines are for $\Thetab=90^\circ$ (diffusion),
and the long-dashed lines are for a blackbody with $5\times 10^6$~K.}
\label{fig:spectrumgeom}
\end{figure*}

Most prominent in the spectra of highly magnetized atmospheres
is the absorption feature at the ion cyclotron resonance
$E_{Bi}=0.63\,(Z/A)B_{14}$~keV. This can be directly traced back to
the corresponding resonance feature in the opacities (see Fig.~\ref{fig:ken}).
The ion cyclotron feature is broad, with $\Delta E/E_{Bi}\sim 1$,
and is much broader than the natural or thermal width of the line
[see Section~\ref{sec:opacity} for a discussion and
eq.~(\ref{eq:ionwidth}) in particular]\footnote{
In discussing the cyclotron feature in this paper, we are
referring to the broad depletion of flux around the ion cyclotron
energy rather than the very narrow ($\Delta E/E\sim 10^{-3}$)
natural/thermal line; such a line has negligible equivalent width
and is not always resolved in
Figs.~\ref{fig:spectrumtemp}-\ref{fig:spectrumcomp}, \ref{fig:spectrumgeom}.
\label{foot:ionwidth}
}.
We find that the ion cyclotron feature is most pronounced when it appears
beyond the blackbody peak, i.e., when $E_{Bi}\ga 3\,k\Teff$.
For the models depicted in Fig.~\ref{fig:spectrum5h}, the
equivalent width of the $E_{Bi}=3.15$~keV line at $B_{14}=5$ is about
1.9~keV, the $0.63$~keV line at $B_{14}=1$ is 90~eV, and
the $29.6$~eV line (not shown) at $B_{12}=4.7$ is only 1.3~eV.

Figure~\ref{fig:spectrumtemp} shows the spectra of H atmospheres at
two different effective temperatures: $\Teff=10^6$~K and $7\times 10^6$~K.
The spectra are similar to those of the $\Teff=5\times 10^6$~K models
shown in Fig.~\ref{fig:spectrum5h}. The cyclotron resonance at
$0.63$~keV (for $B_{14}=1$) becomes prominent for $\Teff=10^6$~K
since the condition $E_{Bi}\ga 3\,k\Teff$ is satisfied.  Note
that the $\Teff=10^6$~K models should be considered for illustrative
purposes only since we expect the opacities to be significantly
modified by neutral atoms and even molecules at such a low temperature
(see Section~\ref{sec:discussion}).

As noted before, in strong magnetic fields, the thermal radiation
is primarily carried by the extraordinary-mode photons. This can be
seen clearly in Figure~\ref{fig:spectrumpolar}. Obviously the
emergent radiation is expected
to be polarized. Also note that the ion cyclotron resonance
feature occurs only for the extraordinary mode, as expected from
the behavior of the opacities (see Sections~\ref{sec:opacity0}
and \ref{sec:opacity}).

Figure~\ref{fig:spectrumcomp} shows the spectra of the He
atmospheres with $\Teff=5\times 10^6$~K, $\Theta_B=0$, and
magnetic fields of $B=0$, $4.7\times 10^{12}$, $10^{14}$
and $5\times 10^{14}$~G. These should be compared to the spectra of
H atmospheres shown in Fig.~\ref{fig:spectrum5h}.  Although
the He atmosphere is denser than the H atmosphere (see
Section~\ref{sec:atmstructure} and Fig.~\ref{fig:tempdens}),
there is no clear distinction between their spectra
except for the ion cyclotron resonance features.
For example, at $B=10^{14}$~G and $\Teff=5\times 10^6$~K,
the flux of the He atmosphere at $E=6$~keV is lower
than that of the H atmosphere by only a factor of 1.2.
Clearly, it would be difficult to distinguish a H atmosphere
from a He atmosphere based on spectral information in the $1-10$~keV
band alone.  We also note that the equivalent widths of the ion
cyclotron features for the helium atmospheres are about 0.51~keV
for the $E_{Bi}=1.58$~keV line at $B_{14}=5$
and about 10~eV for the $E_{Bi}=0.32$~keV line at $B_{14}=1$.

\subsection{Full Radiative Transport Versus Diffusion Approximation}
\label{sec:atmd}

The atmosphere models presented in
Sections~\ref{sec:atmstructure}--\ref{sec:spectra} are based
on solutions of the full radiative transfer equations
(see Sections~\ref{sec:rte} and \ref{sec:numcompfull}).
These models are currently restricted to the case where
the magnetic field is perpendicular to the stellar surface,
i.e., $\Thetab=0$.  Using the diffusion approximation
(see Sections~\ref{sec:rted} and \ref{sec:numcompd}), we
construct atmosphere models for general $\Thetab$'s.

\begin{figure*}
\epsfxsize=15cm
\epsfysize=15cm
\epsffile{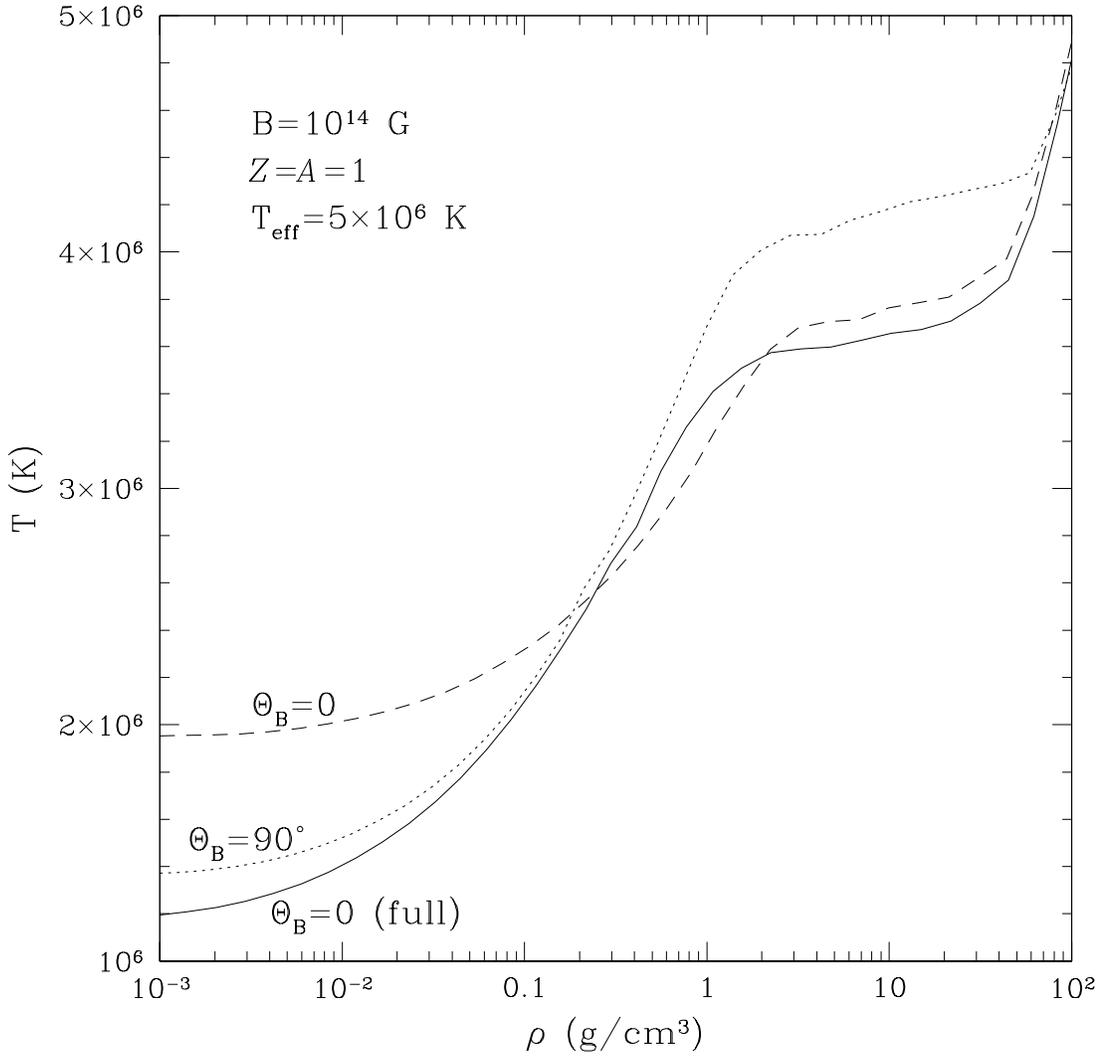}
\caption{Comparison of the temperature-density profiles
between the full radiative transport and diffusion approximated
transport models considered in Section~\ref{sec:atmd}.
The solid line is for $\Thetab=0$ (full radiative transport),
the dashed line is for $\Thetab=0$ (diffusion),
and the dotted line is for $\Thetab=90^\circ$ (diffusion).}
\label{fig:tempdensd}
\end{figure*}

Figure~\ref{fig:tempdensd} shows the temperature-density profiles
of the H atmosphere models with $B=10^{14}$~G, $\Teff=5\times 10^6$~K,
and $\Thetab=0$ and $90^\circ$ based on the diffusion approximation.
These are compared with the result of the corresponding full transport
model with $\Thetab=0$. Clearly, in the diffusion models,
the surface temperature (where $\rho\la 0.01$~g/cm$^3$)
for $\Thetab=90^\circ$ is significantly less than for $\Thetab=0$
(by roughly a factor of 1.4 for the models considered here); this
was already noted by Shibanov \etal~(1992) and Pavlov \etal~(1995) for
the $B\sim 10^{12}$~G models.
At $10^{14}$~G, the temperature plateaus (due to the decoupling of the
extraordinary-mode photons; see Section~\ref{sec:atmstructure})
are also different for $\Thetab=0$ and $\Thetab=90^\circ$.
The difference in the temperature profiles for different magnetic
field geometries implies that different parts of the NS surface
will have different temperatures, e.g., the magnetic equator is
cooler than the magnetic poles, even if $\Teff$ is uniform.
Clearly, a full understanding of the NS surface temperature
distribution requires detailed modeling of the three-dimensional
atmosphere and of the anisotropic heat transport through the
NS crust; this is beyond the scope of this paper.
Comparing the diffusion model with the full transport
model, we see from Figure~\ref{fig:tempdensd} that, at $\Thetab=0$,
the temperature in the full model is significantly lower than
that of the diffusion model in the outermost layers of the atmosphere.

Figure~\ref{fig:spectrumgeom} shows that, despite the difference in
the temperature profiles, the spectra of the diffusion
models for $\Thetab=0$ and $90^\circ$ and the spectra of the
full transport model are rather similar.
For $B_{14}=1$, the equivalent width of the cyclotron feature
at $\Ebi=0.63$~keV is about 80~eV for $\Thetab=0$ and 50~eV
for $\Thetab=90^\circ$, while for $B_{14}=5$, the feature
at $\Ebi=3.15$~keV has a width of about 1.9~keV for $\Thetab=0$ and $90^\circ$.
We also find that the polarized emission is
stronger for $\Thetab=90^\circ$ than for $\Theta_B=0$, as already noted
by Shibanov \etal~(1992) for the $B=4.7\times 10^{12}$~G models.

\setcounter{equation}{0}
\section{Discussion} \label{sec:discussion}

In this paper, we have constructed models of magnetized
neutron star atmospheres composed of ionized hydrogen or helium.
We focused on the superstrong field regime ($B\ga 10^{14}$~G)
as directly relevant to magnetars (see Section~\ref{sec:intro}).
We solve the full, angle-dependent radiative transfer equations for
the coupled polarization modes and include the ion cyclotron and plasma
effects on the opacities and polarization modes. Since the magnetic
field greatly suppresses the opacities of the extraordinary-mode photons,
the thermal radiation emerges from deep in the atmosphere
($\rho\sim 10^2$~g/cm$^3$ for $B=10^{14}$~G), where the medium effect is
non-negligible. As already noted in previous works on neutron star
atmospheres with moderate ($\sim 10^{12}$~G) or negligible magnetic fields,
the thermal emission from a highly magnetized neutron star is polarized,
with the spectrum harder than the blackbody.
We find that the ion cyclotron resonance, at $\Ebi=0.63\,(Z/A)
(B/10^{14}~{\rm G})$~keV, can give rise to a broad ($\Delta E/E
\sim 1$) and potentially observable absorption feature
in the spectrum, especially when $\Ebi\ga 3\,k\Teff$.
Clearly the detection of such a feature would provide
an important diagnostic for magnetars.

We note that the spectra presented in this paper (see
Figs.~\ref{fig:spectrum5h}-\ref{fig:spectrumcomp} and \ref{fig:spectrumgeom})
correspond to emission from a local patch of the neutron star surface.
To directly compare with observations, one must calculate synthetic
spectra from the whole stellar surface, taking into account
the effect of gravitational redshift and light-bending. To do
this, one must know the distribution of the magnetic field
(both magnitude and direction) and the effective temperature
over the stellar surface. For a given field geometry,
the surface temperature distribution may be obtained from crustal
heat conduction calculations
(e.g., Hernquist~1985; Schaaf~1990; Page~1995; Heyl \& Hernquist~1998),
but this is intrinsically
coupled to the properties of the atmosphere; the problem is
further complicated by the existence of lateral radiative flux
(see Section~\ref{sec:rted})
and the nonuniform temperature distribution induced by the magnetic
field in the atmosphere (see Section~\ref{sec:atmd}). Obviously
we expect the cyclotron feature to be broadened and less deep
if different parts of the
surface with different field strengths contribute similarly to the
observed flux (see Zane \etal~2001 for a calculation based on a dipole field
and an approximate temperature distribution).

Several physical effects have been neglected
in our atmosphere models (see \"{O}zel~2001; Zane \etal~2001,
for related works on magnetar atmospheres; see also
Section~\ref{sec:intro} for references of previous works).
We have assumed that the atmosphere is fully ionized.
The problem of ionization equilibrium in a highly magnetized medium is a
complicated one owing to the nontrivial coupling between the center-of-mass
motion and the internal atomic structure and to the
relatively high densities of the atmosphere
(see Lai~2001 and references therein).
While estimates based on the calculations
of Potekhin \etal~(1999) indicate that the fraction of neutral
atoms is no more than a few percent for $T\ga 5\times 10^6$~K at $B\sim
10^{14}$~G, it should be noted that even a small neutral fraction
can potentially affect the radiative opacities. For example,
the ionization edge of a stationary H atom (in the ground state)
occurs at energy $E_i\approx 4.4\,(\ln b)^2$~eV, where $b\equiv B/(2.35\times
10^9~{\rm G})$ (thus, $E_i=0.16,~0.31,~0.54,~0.87$~keV for
$B=10^{12},~10^{13},~10^{14},~10^{15}$~G, respectively; see Lai~2001).
The ratio of the free-free and bound-free opacities
(for the extraordinary mode) at $E=E_i$ is of order
$10^{-6}\rho_1/(T_6^{1/2}B_{14}f_H)$, where
$\rho_1=\rho/(1~{\rm g~cm}^{-3})$, $T_6=T/(10^6~{\rm K})$,
$B_{14}=B/(10^{14}~{\rm G})$, and $f_H$ is the fraction of H atoms in the
``centered'' ground state (see Lai~2001). Thus one could in principle
expect atomic features, broadened by the ``motional Stark effect,''
in the spectra; these could blend with the ion cyclotron feature,
at $E_{Bi}$, studied in this paper.

Another caveat of our models is the neglect of the vacuum polarization
effect. In a strong magnetic field, vacuum polarization
(in which photons are temporarily converted into electron-positron
pairs) contributes to the refractive index by a term of order
$\delta_{\rm vp}=(\alpha_F/45\pi)(B/B_Q)^2$, where $\alpha_F=1/137$ and
$B_Q=4.4\times 10^{13}$~G, and the effect tends to
induce linear polarization of the modes (Adler~1971).
Thus for $3\delta_{\rm vp}\ga (\omega_p/\omega)^2$ (where $\omega_p$ is
the electron plasma frequency), or, for photon energies $\hbar\omega
\ga \hbar\omegavp=(15\pi/\alphaf)^{1/2}(\omegap/\omegab)\me c^2
 = 1.0\,\rho_1^{1/2}B_{14}^{-1}$~keV\footnote{
The expressions of $\delta_{\rm vp}$ and $\hbar\omegavp$ given here
are valid for $B\ll\Bq$.  The use of more general expressions
(Tsai \& Erber~1975; Heyl \& Hernquist~1997b) results in different
scalings with $B$, and the numerical values differ by a factor of
a few for $B\sim 10^{14}-10^{15}$~G (Ho \& Lai~2001). 
}, the normal modes
(and thus the radiative opacities) are modified by vacuum polarization
(e.g., Gnedin \etal~1978, M\'{e}sz\'{a}ros \& Ventura~1979). Previous works
(see M\'{e}sz\'{a}ros~1992 for a review; see also Shibanov \etal~1992,
\"{O}zel~2001 for atmosphere models which
include the vacuum polarization effect; note that \"{O}zel~2001
studied the $B\gg\Bq$ regime but used the vacuum polarization
expressions valid only for $B\ll\Bq$) neglected the response of
the ions in calculating the dielectric tensor. However, as shown in this
paper, for $B\ga 10^{14}$~G, the ion cyclotron resonance gives rise to a
prominent feature in the polarization mode vector and in the opacities.
Therefore one must consider the combined system of electrons, ions,
and polarized vacuum in order to model the spectra in the
superstrong field regime (Ho \& Lai~2001).  Also, vacuum
polarization causes the opacities to exhibit a
resonance feature at $E=\hbar\omegavp$
(Pavlov \& Shibanov~1979; Ventura \etal~1979; see M\'{e}sz\'{a}ros~1992
for a review).
Since $\hbar\omegavp\propto\rho^{1/2}$ and since atmosphere
models necessarily require discrete depth (density) and energy grid
points, such features can appear in the model spectra, while,
in a real atmosphere, they will be smoothed out.
Moreover, vacuum polarization can induce resonant mode conversion
which may significantly affect the radiation transport (Lai \& Ho~2001).
We will present atmosphere models including vacuum
polarization in a future work (Ho \& Lai~2001).

Finally, although in this paper we have included the medium
effect on the polarization modes (see Section~\ref{sec:opacity0}),
at sufficiently high densities or low photon energies, the
refractive index of the medium deviates from unity or
even becomes complex.  This will introduce additional effects
on the radiative transport and the resulting spectra.  This
issue will be addressed in a future work.

\section*{Acknowledgments}

We would like to express our appreciation to Roberto Turolla for his
careful reading of the manuscript and his suggestions and comments.
D.L. thanks Feryal \"{O}zel for discussions of her work
at the ITP conference ``Spin and Magnetism in Young Neutron Stars''
held in October 2000.
This work is supported in part by NASA Grant NAG 5-8484 and
NAG 5-8356 and NSF grant AST 9986740.  D.L. is also supported by
a fellowship from the A.P. Sloan foundation.

\appendix

\setcounter{equation}{0}
\section{Normal Mode Polarization Vector for Magnetized Electron-Ion Plasma}
\label{sec:polarvec}

We outline here the derivation of the polarization vector $\vece^j$
for a magnetized electron-ion plasma.
Many previous works on opacities did not include the ion effects
on the polarization vector (e.g., Ventura~1979; M\'{e}sz\'{a}ros~1992;
see, however, Bulik \& Pavlov~1996).
Our derivation is patterned after Shafranov~(1967)
(see also Ginzburg~1970), who calculated the
index of refraction but did not give the explicit expression
for the polarization vector.

Neglecting the magnetic susceptibility and assuming no external
charges or currents, the Maxwell equations are used to derive
the equation for plane waves with
$\vecE\propto e^{i(\veck\cdot\vecr-\omega t)}$:
\be
\lb\epsilon_{ij}+n^2\lp \hat{k}_i\hat{k}_j-\delta_{ij}\rp\rb E_j=0,
 \label{eq:nrefract}
\ee
where $\hat{k}=\veck/k$, $E_i$ and $k_i$ are the $i$th component
of the electric field and wave vector of the plane wave, respectively,
and $n=ck/\omega$ is the refractive index.
In the frame where the external magnetic field $\vecB$ is
aligned along the $z'$-axis, the dielectric tensor is given by
\be
\lb\epsilon_{ij}'\rb = \lb \begin{array}{ccc}
\varepsilon & ig & 0 \\
-ig & \varepsilon & 0 \\
0 & 0 & \eta
\end{array} \rb, \label{eq:epsij0}
\ee
where
\ba
\varepsilon & = &
 1-\sum_s\frac{\lambdad v}{\lambdad^2 - u} \label{eq:epsii0} \\
\eta & = &
 1-\sum_s\frac{v}{\lambdad} \label{eq:epszz0} \\
g & = &
 -\sum_s\frac{u^{1/2}v}{\lambdad^2-u}. \label{eq:epsxy0}
\ea
Here the sums run over each charged particle species $s$
in the plasma,
and $u = \omegabo^2/\omega^2$ and $v = \omegapo^2/\omega^2$, where
$\omegabo$ and $\omegapo$ are the cyclotron and plasma frequencies
of charged particle $s$, respectively.
Damping of the particle motion is accounted for in
$\lambdad = 1+i\nu_s/\omega$, where $\nu_s$ is the damping rate of
a particle of species $s$.
As in previous studies, damping is usually taken to be small, so that
$\lambdad\rightarrow 1$.
In the frame where the propagation wave vector $\veck$ is aligned
along the $z$-axis and $\vecB$ is in the $x$-$z$ plane, the
dielectric tensor is given by
\be
\lb\epsilon_{ij}\rb = \lb \begin{array}{ccc}
\varepsilon\cos^2\thetab+\eta\sin^2\thetab & ig\cos\thetab &
 \lp\varepsilon-\eta\rp\sin\thetab\cos\thetab \\
-ig\cos\thetab & \varepsilon & -ig\sin\thetab \\
\lp\varepsilon-\eta\rp\sin\thetab\cos\thetab & ig\sin\thetab &
 \varepsilon\sin^2\thetab+\eta\cos^2\thetab
\end{array} \rb. \label{eq:epsijx}
\ee
Substituting equation~(\ref{eq:epsijx}) into equation~(\ref{eq:nrefract})
and setting the determinant of the resulting equation to zero
yields the index of refraction,
\be
n^2 = \frac{B \pm \lp B^2 - 4AC\rp^{1/2}}{2A}, \label{eq:nrefract3}
\ee
where
\ba
A & = & \varepsilon\sin^2\thetab+\eta\cos^2\thetab \\
B & = & \lp\varepsilon^2-g^2-\varepsilon\eta\rp\sin^2\thetab+2\varepsilon\eta \\
C & = & \lp\varepsilon^2-g^2\rp\eta.
\ea

To obtain the polarization vector, we write the electric
field as $\vecE=\vece^j=e_0(iK_j,1,iK_{z,j})$, where
$iK_j=E_x/E_y$, $iK_{z,j}=E_z/E_y$, and $e_0=(1+K_j^2+K_{z,j}^2)^{-1/2}$
from the normalization $|\vecE|^2=1$.
From equation~(\ref{eq:nrefract}), we find
\be
K_{z,j}=-\epsilon_{zz}^{-1}\lp\epsilon_{zx}K_j-i\epsilon_{zy}\rp,
 \label{eq:polarkz0}
\ee
which gives equation~(\ref{eq:polarkz}) when considering the case
of electrons and one species of ions.
Equation~(\ref{eq:nrefract}) can also be used to obtain $K_j$, which is
given by equation~(\ref{eq:polarkj}), with
\be
\polarb = -\frac{\lp\varepsilon^2-g^2-\varepsilon\eta\rp
 \sin^2\thetab}{2g\eta\cos\thetab}. \label{eq:polarb0}
\ee
Specializing to the case of electrons and one species of ions,
equation~(\ref{eq:polarb0}) can be simplified to equation~(\ref{eq:polarb}).
Once $\vece^j$ is rotated into the frame with $\vecB$ along the $z$-axis,
the polarization vector becomes equations~(\ref{eq:polarvecpm})
and (\ref{eq:polarvecz}).


\label{lastpage}

\clearpage

\setcounter{section}{0}
\clearpage

\clearpage

\clearpage

\clearpage

\clearpage

\clearpage

\clearpage

\clearpage

\clearpage

\clearpage

\clearpage

\end{document}